\documentclass[twocolumn,epjc3]{svjour3mod}

\usepackage{amsmath,amssymb,amsfonts,dcolumn,mathcomp,slashed,dsfont}
\usepackage{graphicx}
\usepackage{bm}
\usepackage{xcolor}
\usepackage{csquotes}
\usepackage{units}
\usepackage{flushend}
\usepackage{cite}
\usepackage[colorlinks,citecolor=blue,urlcolor=blue,linkcolor=blue]{hyperref}
\usepackage{ragged2e}
\usepackage{comment}
\usepackage[compat=1.1.0]{tikz-feynman}
\usetikzlibrary{shapes,arrows,positioning,automata,backgrounds,calc,er,patterns}
\usepackage{siunitx}
\usepackage{subcaption}
\usepackage{booktabs}

\apptocmd\thebibliography\justifying{}{}

\newcommand{\Tr}{\text{Tr}}
\newcommand{\Omnes}{Omnès }
\newcommand{\M}{\mathcal{M}}

\newcommand{\F}{\mathcal{F}}
\newcommand{\G}{\mathcal{G}}
\renewcommand{\L}{\mathcal{L}}
\newcommand{\Ha}{\mathcal{H}}

\newcommand{\Fpi}{F_\pi}
\newcommand{\Order}{\mathcal{O}}
\newcommand{\disc}{{\rm disc}\,}
\newcommand{\dd}{{\rm d}}
\newcommand{\mpi}{M_{\pi}}

\newcommand{\mK}{M_{K}}

\newcommand{\sth}{s_\text{th}}
\newcommand{\sKs}{s_{K^*}}

\renewcommand{\Re}{{\rm Re}\,}
\renewcommand{\Im}{{\rm Im}\,}

\newcommand{\gn}{d^{\text{n}}_{K^*}}
\newcommand{\gc}{d^{\text{c}}_{K^*}}

\newcommand{\GeV}{\,\text{GeV}}
\newcommand{\MeV}{\,\text{MeV}}
\newcommand{\bsp}{\begin{sloppypar}}
\newcommand{\esp}{\end{sloppypar}}

\newcommand*{\tausche}[2]{
	\expandafter\mathchardef\expandafter#2\number\expandafter#1\expandafter\relax
	\expandafter\mathchardef\expandafter#1\number#2}
\tausche{\epsilon}{\varepsilon}

\graphicspath{{./plots/amplitudes/}{./plots/cross_sections/}{./plots/figures/}}

\begin{document}

\title{\boldmath Dispersive analysis of the Primakoff reaction $\gamma K \to K \pi $}

\author{Maximilian Dax\thanksref{addr,e1}
\and
Dominik Stamen\thanksref{addr,e2} 
\and
Bastian Kubis\thanksref{addr,e3}
}

\thankstext{e1}{e-mail: dax@hiskp.uni-bonn.de}
\thankstext{e2}{e-mail: stamen@hiskp.uni-bonn.de}
\thankstext{e3}{e-mail: kubis@hiskp.uni-bonn.de}

\institute{Helmholtz-Institut f\"ur Strahlen- und Kernphysik (Theorie) and
   Bethe Center for Theoretical Physics,
   Universit\"at Bonn, 
   53115 Bonn, Germany \label{addr}
}

\date{}

\maketitle

\begin{abstract}
\bsp
We provide a dispersion-theoretical representation of the reaction amplitudes $\gamma K\rightarrow K \pi$ in all charge channels, based on modern pion--kaon $P$-wave phase shift input.  
Crossed-channel singularities are fixed from phenomenology as far as possible.
We demonstrate how the subtraction constants can be matched to a low-energy theorem and radiative couplings of the $K^*(892)$ resonances,
thereby providing a model-independent framework for future analyses of high-precision kaon Primakoff data. 
\esp
\end{abstract}

\section{Introduction}
The Wess--Zumino--Witten anomaly~\cite{Wess:1971yu,Witten:1983tw} provides QCD predictions for processes of odd intrinsic parity at low energies.  The textbook example is the two-photon decay of the neutral pion~\cite{Adler:1969gk,Bell:1969ts,Bardeen:1969md}, which is determined, at zero quark masses, by the elementary charge, $e$, and the pion decay constant, $F_\pi$.  The next-more-complicated reactions involving strong and electromagnetic interactions only are three-pseudoscalars--photon processes~\cite{Bijnens:1989ff} such as $\gamma\pi\to\pi\pi$, or $\eta\to\pi\pi\gamma$.  Low-energy theorems for these are of a very similar structure, i.e., they provide parameter-free predictions in terms of $e$ and $F_\pi$, e.g.
\begin{equation}
    F_{3\pi} = \frac{e}{4\pi^2F_\pi^3}
    \label{eq:3pi-anomaly}
\end{equation}
for $\gamma\pi\to\pi\pi$~\cite{Adler:1971nq,Terentev:1971cso,Aviv:1971hq}.  
This reaction can be investigated experimentally in a Primakoff reaction~\cite{Primakoff:1951pj}, with a charged-pion beam scattered off the Coulomb field of a heavy nucleus.  Such experiments have been performed, with the objective to test the prediction of Eq.~\eqref{eq:3pi-anomaly}, at Serpukhov~\cite{Antipov:1986tp}, or are being analyzed at COMPASS~\cite{Seyfried2017}.  

The generalization of such Primakoff reactions to beams of charged kaons was conceived as early as in the 1960s~\cite{Halprin:1966zz}, and put into practice in the 1970s both at the CERN Proton Synchrotron~\cite{Bemporad:1973pa} and at AGS in Brookhaven~\cite{Carithers:1975cg}, with refined experiments conducted in the 1980s at Fermilab~\cite{Berg:1980cp,Cihangir:1982ze,Chandlee:1983hf,Carlsmith:1985ep}.  The motivation here was mainly the exploration of radiative couplings of strange resonances, predominantly the $K^*(892)$, and the supposed relation of these (magnetic) radiative transitions, in the quark model, to quark magnetic moments~\cite{Berg:1980cp,Chandlee:1983hf}.  KTeV has investigated Primakoff production of neutral strange resonances with a $K_L$ beam~\cite{AlaviHarati:2001ic}, concentrating on radiative widths of heavier kaon resonances.  Currently, the OKA experiment~\cite{Burtovoy:2016dyf,Burtovoy:2017hbt} analyzes data on charged-kaon Primakoff reactions.  In the future, high-precision data is expected from the upgrade to a kaon beam at COMPASS++/AMBER~\cite{Abbon:2014aex,Denisov:2018unj,Bernhard:2019jqz}.

It was realized in Refs.~\cite{Hoferichter:2012pm,Hoferichter:2017ftn} for the anomalous photon--pion reaction that both aspects, low-energy theorem and chiral anomaly on the one hand, and radiative resonance couplings on the other, are intimately related to each other.
Unitarity implies a close link between the amplitude $\gamma\pi\to\pi\pi$, at zero energy and in the chiral limit, and its behavior in the resonance peak region of the $\rho(770)$.  This has the practical consequence that the prediction due to the anomaly can be tested with much better statistics~\cite{Hoferichter:2012pm}.  In addition, using a dispersion-theoretical representation, the radiative coupling $\rho\to\pi\gamma$ can be extracted in a model-independent way, from the residue of the pole on the second Riemann sheet~\cite{Hoferichter:2017ftn}.
Furthermore, such dispersive amplitudes will also help to link lattice QCD calculations~\cite{Briceno:2015dca,Briceno:2016kkp,Alexandrou:2018jbt} to physical parameters~\cite{Niehus:2019nkl}.  The overarching interest in $\gamma\pi\to\pi\pi$ is also justified by its role in the dispersive reconstruction of the neutral-pion transition form factor~\cite{Hoferichter:2014vra,Hoferichter:2018dmo,Hoferichter:2018kwz} and the latter's role for hadronic light-by-light scattering and the muon's anomalous magnetic moment~\cite{Colangelo:2014pva,Aoyama:2020ynm}.

In this article, we construct a dispersion-theoretical representation for $\gamma K\to K\pi$ (in all possible charge configurations) that fulfills a similar feat.  The chiral anomaly predicts the amplitudes for $\pi^0$ production to have the exact same value in the chiral limit and at zero energy as the analogous photon--pion reaction, see Eq.~\eqref{eq:3pi-anomaly}; based on the fundamental principles of analyticity and unitarity, the anomaly can also here be related to the radiative couplings of $K^*(892)\to K \gamma$~\cite{Berg1983,Chandlee:1983hf,Carlsmith:1985ep}.
In this manner, our analysis provides a consistent framework to analyze future data, from OKA or COMPASS++/AMBER, in a theoretically sound setting.  The radiative $K^*$ coupling constants are, inter alia, important input quantities for coupled-channel descriptions of photon--photon fusion reactions $\gamma\gamma\to \pi\pi/K\bar K$~\cite{GarciaMartin:2010cw,Danilkin:2019opj} and $\gamma\gamma\to\pi\eta/K\bar K$~\cite{Danilkin:2017lyn,Lu:2020qeo}. 
On a similar note, the full $\gamma K\to K\pi$ amplitudes might serve as a building block for an advanced analysis of Compton scattering on kaons, from which the kaon polarizabilities~\cite{Guerrero:1997rd} can be extracted, the main motivation of the COMPASS kaon Primakoff program~\cite{Moinester:2003rb}.

Previous theoretical work on these reactions is rather elusive.  The channels with incoming charged kaons were calculated in Ref.~\cite{Vysotsky:2015mks} in a tree-level model based on effective Lagrangians for vector exchanges. 
For $\gamma K^-\to K^-\pi^0$ only, one-loop corrections in the chiral expansion have been considered~\cite{Ebertshaeuser2001,Hacker2008}. 
Here we derive Khuri--Treiman-type equations~\cite{Khuri:1960zz} for all possible charge configurations and solve these self-consistently for the (crossing-symmetric) $s$- and $u$-channels, while $t$-channel singularities are fixed from data and symmetry arguments as much as possible.  To guarantee an accurate description of the universal kaon--pion final-state interactions, we employ phase shift input from corresponding Roy--Steiner analyses~\cite{Buettiker:2003pp,Pelaez:2016tgi,Pelaez:2020gnd}.

The outline of this article is as follows. 
We introduce the necessary kinematics as well as partial-wave and isospin formalism in Sec.~\ref{sec:kinematics-and-amplitude}, leading to the amplitude decomposition in terms of so-called reconstruction theorems.
The fixed $t$-channel amplitudes are determined in Sec.~\ref{sec:tchannel}.
Section~\ref{sec:disp} is devoted to derivation and solution of the Khuri--Treiman equations for $\gamma K\to K\pi$.
In Sec.~\ref{sec:matching}, we discuss the matching of the subtraction constants, the free parameters of the dispersive representation, to the chiral anomaly and the radiative $K^*$ couplings.
Results for partial waves and cross sections are shown in Sec.~\ref{sec:results}.
We summarize and conclude our study in Sec.~\ref{sec:summary}.
Some technical aspects about dispersive kernel functions are relegated to an appendix.

\section{Decomposition of the amplitude}\label{sec:kinematics-and-amplitude}
\subsection{Kinematics and partial-wave decomposition}
We decompose the amplitude for the reaction
\begin{equation}
\label{eq:definition-momenta-scattering-process}
\gamma(q)K(p_1)\rightarrow K(p_2) \pi(p_0)
\end{equation}
in terms of a kinematic prefactor of odd intrinsic parity and the scalar amplitude $\F(s,t,u)$ according to
\begin{equation}
\label{eq:decomposition-full-amplitude-scalar-amplitude}
\M=i\epsilon_{\mu\nu\alpha\beta}\varepsilon^\mu p_1^\nu p_2^\alpha p_0^\beta \F(s,t,u),
\end{equation}
where $\varepsilon^\mu$ is the polarization vector of the photon and the Mandelstam variables~\cite{Mandelstam:1958xc} are given by $s=\left(q+p_1\right)^2$, $t=\left(q-p_0\right)^2$, and $u=\left(q-p_2\right)^2$. In the isospin limit with $M_\pi=M_{\pi^\pm}$ and $M_K=0.496\GeV$ (the convention used in the pion--kaon scattering analysis of Ref.~\cite{Pelaez:2016tgi}), the on-shell condition reads $s+t+u=2M_K^2+M_\pi^2=3s_0$ and the elastic threshold is given by $\sth = \left(M_K+M_\pi\right)^2$. 
The cosine of the $s$-channel center-of-mass scattering angle $z_s=\cos\theta_s$ and the two-body phase space factor $\kappa(s)$ are defined by
\begin{align}
z_s&=\frac{s(t-u)-\mK^2\left(\mpi^2-\mK^2\right)}{s\left(s-\mK^2\right)\kappa(s)}\,,\notag\\
\kappa(s)&=\frac{\lambda^{1/2}(s,\mpi^2,\mK^2)}{s}\,,
\end{align}
with the K\"all\'en function $\lambda(x,y,z) = x^2+y^2+z^2-2(xy+xz+yz)$.
The Mandelstam variables $t$ and $u$ can be expressed in terms of $s$ and $z_s$ according to
\begin{align}
t &= a^{+\Delta}(s)+b(s)z_s\,, &
a^{\pm \Delta}(s) &= \frac{3s_0-s}{2}\mp\frac{\Delta}{2s}\,, \notag\\
u &= a^{-\Delta}(s)-b(s)z_s\,, &
b(s)&= \frac{\left(s-\mK^2\right)\kappa(s)}{2}\,, \notag\\
\Delta &= \mK^2\left(\mpi^2-\mK^2\right)\,. 
\label{eq:a-b-def}
\end{align}
The partial-wave decomposition and its inversion are given by~\cite{Jacob:1959at}
\begin{align}
\F(s,t,u)&=\sum_{\ell}f_\ell(s)P'_\ell(z_s)\label{eq:defdecomp}\,,\\
f_\ell(s)&=\frac{1}{2}\int_{-1}^{1}\dd{z_s} \big[P_{\ell-1}(z_s)-P_{\ell+1}(z_s)\big] \F(s,t,u)\,, \notag
\end{align}
in terms of derivatives of the Legendre polynomials.
In particular, the $P$-wave that dominates at low energies is found to be
\begin{equation}
    f_1(s)=\frac{3}{4}\int_{-1}^{1}\dd{z_s} (1-z_s^2)\ \F(s,t,u)\,.
    \label{eq:Pwave-projection}
\end{equation}
Finally, the total cross section for $\gamma K\to K\pi$ is given by
\begin{align}
\sigma(s)&=\frac{(s-M_K^2)\lambda^{3/2}(s,M_\pi^2,M_K^2)}{1024\pi s^2} \notag\\
& \quad \times \int_{-1}^{1}\dd z_s (1-z_s^2)|\F(s,t,u)|^2 \notag\\
&=\frac{(s-M_K^2)\lambda^{3/2}(s,M_\pi^2,M_K^2)}{768\pi s^2} \left(|f_1(s)|^2 + \ldots\right) \,,\label{eq:crosssection}
\end{align}
where the ellipsis denotes $D$- and higher partial waves.

In the following, we will also discuss the corresponding $t$-channel reaction
\begin{equation}
\label{eq:definition-momenta-scattering-process-t-channel}
\gamma(q)\pi(-p_0)\rightarrow K(p_2) \bar{K}(-p_1) \,,
\end{equation}
for which the partial-wave expansion is defined analogously to Eq.~\eqref{eq:defdecomp} with 
the cosine of the $t$-channel scattering angle $z_t$ 
\begin{equation}
z_t = \frac{t(s-u)}{s(s-\mpi^2)\sigma_K(s)}\,, ~~
\text{where} ~~
    \sigma_P(s)=\sqrt{1-\frac{4M_P^2}{s}} 
\end{equation}
is the two-body phase space for equal masses.
We will analyze the $t$-channel
in terms of intermediate $\pi\pi$ states, for which the photon--pion amplitude $\F_{\gamma\pi\to\pi\pi}(s,t,u)$ defined analogously to Eq.~\eqref{eq:decomposition-full-amplitude-scalar-amplitude} serves as an input.
We denote the corresponding partial waves (which necessarily have isospin $I=1$) by $h_\ell^1(t)$, 
\begin{equation}
   \F_{\gamma\pi\to\pi\pi}(s,t,u) =\sum_{\ell~\textrm{odd}}h_\ell^1(t)P'_\ell(z_t)\,.
   \label{eq:PWE-gamma3pi}
\end{equation}

Finally, we need to introduce the partial waves for two meson--meson scattering amplitudes. In the $s$-channel, pion--kaon elastic scattering of isospin $I$ is expanded in partial waves $t_\ell^I(s)$ according to
\begin{equation}
    T_{K\pi\to K\pi}^I(s,t,u) = 16\pi\sum_{\ell=0}^\infty (2\ell+1)P_\ell(z_s)t_\ell^I(s) \,,
\end{equation}
which are parameterized in terms of the scattering phase shifts $\delta_\ell^I(s)$ according to
\begin{equation}
t^I_\ell(s)=\frac{e^{i\delta_\ell^I(s)}\sin\delta_\ell^I(s)}{\kappa(s)}\,.
\end{equation}
In the $t$-channel, we employ a slightly different convention for the $\pi\pi\rightarrow K\bar{K}$ partial waves $g_\ell^I(t)$ to keep them free of kinematical singularities,
\begin{align}
\label{eq:PWE-t-channel}
&T_{\pi\pi\rightarrow K\bar{K}}^I(s,t,u) \notag\\
&=16\pi \sqrt{2}\sum_{\ell=0}^{\infty} (2\ell+1) \big[q_\pi(t) q_K(t)\big]^\ell P_\ell(z_t)g_\ell^I(t)\,, 
\end{align}
where $q_P(t) = \sqrt{t}\,\sigma_P(t)/2$.

\subsection{Isospin decomposition and reconstruction theorems}\label{sec:isospin}

In terms of isospin, the reaction $\gamma K\rightarrow K\pi$ is equivalent to pion photoproduction off a nucleon $\gamma N\rightarrow N\pi$ studied, e.g., in
Refs.~\cite{Chew:1957tf,Tiator:1990ck,Drechsel:1992pn,Hanstein:1996bd,Hanstein:1997tp,Pasquini:2006yi}.  
The isospin decomposition for the latter process thus also applies here:
\begin{align}
\F^{-0}\big(\gamma K^-\rightarrow K^-\pi^0\big)&=A^{(+)}-A^{(0)}\,,\notag\\
\F^{0-}\big(\gamma K^-\rightarrow \bar{K}^0\pi^-\big)&=-\sqrt{2}\left(A^{(-)}-A^{(0)}\right)\,,\notag\\
\F^{00}\big(\gamma\bar{K}^0\rightarrow\bar{K}^0\pi^0\big)&=A^{(+)}+A^{(0)}\,,\notag\\
\F^{-+}\big(\gamma \bar{K}^0\rightarrow K^-\pi^+\big)&=\sqrt{2}\left(A^{(-)}+A^{(0)}\right)\,.
\label{eq:isospin-decomp-t1}
\end{align}
The amplitude $A^{(0)}$ is associated with the isoscalar photon component. It has total isospin $I=1/2$ in the $s$- and $u$-channels and isospin $I=1$ in the $t$-channel. 
The amplitudes $A^{(+)}$ and $A^{(-)}$ on the other hand correspond to an isovector photon. In the $t$-channel, they have pure isospin $I=0$ and $I=1$, respectively. 
$A^{(+)}$ and $A^{(-)}$ can be decomposed in terms of amplitudes $A^{(1/2)}$ and $A^{(3/2)}$ of definite isospin $I=1/2$ and $I=3/2$ in the $s$- and $u$-channels according to
\begin{align}
A^{(+)}&=\frac{1}{3}\left(A^{(1/2)}+2A^{(3/2)}\right)\,,\notag\\
A^{(-)}&=\frac{1}{3}\left(A^{(1/2)}-A^{(3/2)}\right)\,.
\end{align}

For dispersion-theoretical analyses of scattering or (three-body) decay amplitudes, it is highly advantageous to decompose these in terms of single-variable amplitudes (SVAs).
Decompositions of such a kind are commonly referred to as reconstruction theorems~\cite{Stern:1993rg,Knecht:1995tr,Zdrahal:2008bd}. 
While the end results have the simple appearance of a combined partial-wave expansion
simultaneously in all three Mandelstam variables,
the derivation requires a careful analysis of the amplitudes using fixed-$s$, -$t$, and -$u$ dispersion relations~\cite{Dax2020}.
With one exception, we neglect discontinuities of partial waves with $\ell\geq 2$, 
resulting in the following reconstruction theorems for
$A^{(0)}$, $A^{(+)}$, and $A^{(-)}$:
\begin{align}
    \label{eq:RT}
    A^{(0)}(s,t,u) &=\F^{(0)}(s)+\G^{(0)}(t)+\F^{(0)}(u)\,, \notag\\
    A^{(+)}(s,t,u)&=\F^{(1/2)}(s) + \F^{(3/2)}(s) + \G^{(+)}(t) \notag\\ & + \F^{(1/2)}(u) + \F^{(3/2)}(u) \,, \notag\\
    A^{(-)}(s,t,u)&=\F^{(1/2)}(s) - \frac{1}{2} \F^{(3/2)}(s) + (u-s) \Ha^{(-)}(t) \notag\\ 
    & - \F^{(1/2)}(u) + \frac{1}{2} \F^{(3/2)}(u) \,.
\end{align}
It can be shown that the decomposition~\eqref{eq:RT}, omitting the $D$-wave SVA $\Ha^{(-)}(t)$, is sufficient to reproduce the chiral expansion of the $\gamma K\to K\pi$ amplitudes exactly up to corrections of $\Order(p^{10})$.
We retain $\Ha^{(-)}(t)$  since the corresponding $P$-wave is forbidden by charge conjugation, such that the $D$-wave SVA represents the leading $t$-channel discontinuity
in $A^{(-)}$.  
Moreover, $\Ha^{(-)}$ is enhanced due to the resonant $a_2(1320)$ contribution.  

Despite the (potentially) very high accuracy of the representation~\eqref{eq:RT} at low energies, the range of applicability towards higher energies is clearly limited.  One of the main limiting factors for a description of cross-section data in the direct or $s$-channel is the appearance of a resonant $K\pi$ $D$-wave around the $K_2^*(1430)$, which has been measured to also couple to $\gamma K$ at least in the charged case~\cite{Cihangir:1982ze}.  Given a width of $\Gamma_{K_2^*(1430)} = 100(2)\MeV$, this suggests our representation to be applicable up to well below $\sqrt{s} = 1.3\GeV$.  Furthermore, for the dominant $K\pi$ $P$-wave, we will stick to the implementation of elastic unitarity with $K\pi$ intermediate states only, which will break down around the $K^*(1410)$ resonance ($\Gamma_{K^*(1410)} = 232(21)\MeV$) with its large inelastic coupling mainly to $K\pi\pi$.\footnote{See, e.g., Refs.~\cite{Moussallam:2007qc,Bernard:2013jxa} for a coupled-channel treatment of $K\pi \leftrightarrow K\pi\pi$ in the $P$-wave.} For this reason, the dispersive amplitude representation we aim for is supposed to be valid to good approximation up to $\sqrt{s} = 1.2\GeV$. As a side remark, we point out that our analysis is at this point limited to real photons, and hence its isovector part is not immediately transferable to the vector current in corresponding $\tau$ decays, $\tau^-\to (K\bar K\pi)^-\nu_\tau$.

The quantum numbers of all aforementioned SVAs and the corresponding dominant resonances in the low-energy regime are summarized in Table~\ref{tab:isospin-SVAs}. 

The decompositions in Eq.~\eqref{eq:RT} are not unique, as only the full amplitudes are physically observable, not the individual SVAs.  It is easily checked that $A^{(0)}$, $A^{(+)}$, and $A^{(-)}$ are invariant under the following shifts in the SVAs: 
\begin{align}
    \delta\F^{(0)}(s) &= \alpha + \delta (s-s_0) \,, \notag\\
    \delta\G^{(0)}(t) &= -2\alpha + \delta (t-s_0)\,, 
    \notag\\
    \delta\F^{(1/2)}(s) &= \beta - \gamma + 2\epsilon(s-s_0)-2\eta\, s \,, 
    \notag\\
    \delta\F^{(3/2)}(s) &= -\beta-\gamma+2\epsilon(s-s_0)+2\eta\, s\,, \notag \\
    \delta\G^{(+)}(t) &= 4\gamma +4\epsilon (t-s_0) \,, \notag\\
    \delta\Ha^{(-)}(t) &= \epsilon-3\eta\,. \label{eq:ambiguity}
\end{align}
In the following, we omit non-constant polynomial shifts from consideration, as we will constrain the SVAs not to grow for large arguments; this implies $\delta=\epsilon=\eta=0$. Therefore these shifts are forbidden if and only if the SVAs satisfy the asymptotic constraints.
We will furthermore fix the $t$-channel SVAs, such that the only remaining ambiguity in Eq.~\eqref{eq:ambiguity} is due to $\beta$, which allows us to shift a constant between $\F^{(1/2)}$ and $\F^{(3/2)}$.

\begin{table}
\renewcommand{\arraystretch}{1.3}
	\centering
	\begin{tabular}{cccccccc}
        \toprule
		&$I_s$& $I_t$& $I_u$& $I_\gamma$&$\eta_G$&$\eta_C$&Resonance\\
		\midrule
		\rule{0pt}{12pt}$\F^{(1/2)}$&$1/2$&/&$1/2$&$1$&/&/&$K^*$\\
		\rule{0pt}{12pt}$\F^{(3/2)}$&$3/2$&/&$3/2$&$1$&/&/&\\
		\rule{0pt}{12pt}$\F^{(0)}$&$1/2$&/&$1/2$&$0$&/&/&$K^*$\\
		\rule{0pt}{12pt}$\G^{(0)}$&/&$1$&/&$0$&$+$&$-$&$\rho$\\
		\rule{0pt}{12pt}$\G^{(+)}$&/&$0$&/&$1$&$-$&$-$&$\omega,\phi$\\
		\rule{0pt}{12pt}$\Ha^{(-)}$&/&$1$&/&$1$&$-$&$+$&$a_2$\\
		\bottomrule
	\end{tabular}
	\caption{
		Quantum numbers of the six single-variable amplitudes. The isospin in the $s$-, $t$-, and $u$-channels is denoted by $I_{s}$, $I_{t}$, and $I_{u}$, respectively. $I_\gamma$ refers to the isospin of the photon. The $C$ and $G$ eigenvalues of the single-variable amplitudes are denoted by $\eta_C$ and $\eta_G$. The `$/$' symbol indicates a non-definite value. In the last column the resonances that dominate the respective SVAs are given.
	}
	\label{tab:isospin-SVAs}
\renewcommand{\arraystretch}{1.0}
\end{table}

\section{Singularities in the $t$-channel}\label{sec:tchannel}

The usual approach to analyzing Khuri--Treiman-type systems is to solve the unitarity relations for the single-variable amplitudes in all three channels fully self-consistently. This is an obvious strategy for perfectly crossing-symmetric systems such as $\gamma\pi\to\pi\pi$~\cite{Hoferichter:2012pm}, related three-pion decays~\cite{Niecknig:2012sj}, or even pion--pion scattering~\cite{Albaladejo:2018gif}, but has also been followed for less symmetric processes such as $\eta\to\pi^+\pi^-\pi^0$~\cite{Colangelo:2018jxw}, $\eta'\to\eta\pi\pi$~\cite{Isken:2017dkw}, or $D\to \bar K\pi\pi$~\cite{Niecknig:2015ija,Niecknig:2017ylb}.  
We do not pursue this approach here as far as the $t$-channel is concerned, for the following reasons: the $t$-channel singularities in $\gamma K\to K\pi$ are either dominantly inelastic ($\rho$, $a_2$), or consist of very narrow poles ($\phi$), or both ($\omega$). For this reason, in our analysis we approximate these by fixed $t$-channel contributions, similar in spirit e.g.\ to various analyses of $\gamma\gamma\to\pi\pi$~\cite{GarciaMartin:2010cw,Hoferichter:2011wk} or the description of left-hand cuts in $\eta^{(\prime)}\to\pi^+\pi^-\gamma$~\cite{Kubis:2015sga}.

\subsection{$\G^{(+)}$ using vector meson dominance}
\label{sec:VMD}

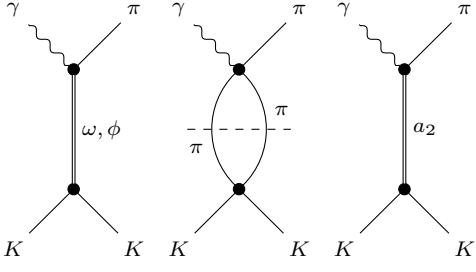
\begin{figure}[t]
	\centering
	\begin{tikzpicture}[baseline=0ex,scale=1]
	\begin{feynman}
	\vertex(a1) {$\gamma$};
	\vertex[right=5em of a1](a2) {$\pi$};
	\vertex[right=2.5em of a1](a3) {};
	\vertex[below=2.5em of a3,dot](b1) {};
	\vertex[below=5em of b1,dot](b2) {};
	\vertex[below=2.5em of b2](c3) {};
	\vertex[left=2.5em of c3](c1) {$K$};
	\vertex[right=2.5em of c3](c2) {$K$};
	\vertex[below=5em of a1](d1) {};
	\vertex[below=5em of a2](d2) {};
	\diagram* {
		(a1)--[photon](b1),(a2)--(b1),(b1)--[double, edge label={$\omega,\phi$}](b2),(b2)--(c1),(b2)--(c2)
	};
	\end{feynman}
	\end{tikzpicture}
		\begin{tikzpicture}[baseline=0ex,scale=1]
	\begin{feynman}
	\vertex(a1) {$\gamma$};
	\vertex[right=5em of a1](a2) {$\pi$};
	\vertex[right=2.5em of a1](a3) {};
	\vertex[below=2.5em of a3,dot](b1) {};
	\vertex[below=5em of b1,dot](b2) {};
	\vertex[below=2.5em of b2](c3) {};
	\vertex[left=2.5em of c3](c1) {$K$};
	\vertex[right=2.5em of c3](c2) {$K$};
	\vertex[below=5em of a1](d1) {};
	\vertex[below=5em of a2](d2) {};
	\diagram* {
		(a1)--[photon](b1),(a2)--(b1),(b1)--[bend left=45, edge label=$\pi$, pos=0.45](b2),(b2)--[bend left=45, edge label=$\pi$, pos=0.45](b1),(b2)--(c1),(b2)--(c2),(d1)--[scalar](d2)
	};
	\end{feynman}
	\end{tikzpicture}
	\begin{tikzpicture}[baseline=0ex,scale=1]
	\begin{feynman}
	\vertex(a1) {$\gamma$};
	\vertex[right=5em of a1](a2) {$\pi$};
	\vertex[right=2.5em of a1](a3) {};
	\vertex[below=2.5em of a3,dot](b1) {};
	\vertex[below=5em of b1,dot](b2) {};
	\vertex[below=2.5em of b2](c3) {};
	\vertex[left=2.5em of c3](c1) {$K$};
	\vertex[right=2.5em of c3](c2) {$K$};
	\vertex[below=5em of a1](d1) {};
	\vertex[below=5em of a2](d2) {};
	\diagram* {
		(a1)--[photon](b1),(a2)--(b1),(b1)--[double, edge label={$a_2$}](b2),(b2)--(c1),(b2)--(c2)
	};
	\end{feynman}
	\end{tikzpicture}
	\caption{$t$-channel contributions to $\gamma K\rightarrow K\pi$; see main text for the individual terms.}
	\label{fig:t-channel-diagram}
\end{figure}

\begin{sloppypar}
The isospin $I=0$ $P$-wave amplitude $\G^{(+)}(t)$ is dominated by the $\omega$ and $\phi$ vector mesons. We calculate this SVA via tree-level exchanges with intermediate vector resonances $V\in\{\omega,\phi\}$, see Fig.~\ref{fig:t-channel-diagram}. To this end we use the effective Lagrangian from Ref.~\cite{Klingl:1996by}. Encoding the relevant Goldstone bosons in
\begin{equation}
\label{eq:Goldstone-boson-matrix-Phi}
\Phi = \sqrt{2}\left(
\begin{array}{ccc}
\frac{\pi^0}{\sqrt{2}} & \pi^+ & K^+\\
\pi^- & \frac{-\pi^0}{\sqrt{2}} & K^0\\
K^- & 
\bar{K}^0 
& 0
\end{array}
\right)
\end{equation}
and the vector mesons in
\begin{equation}
V_\mu=\sqrt{2}\left(\begin{array}{ccc}
\frac{\rho+\omega}{\sqrt{2}} & \rho^+ &K^{*+} \\
\rho^- &\frac{-\rho+\omega}{\sqrt{2}}  &K^{*0} \\
K^{*-}& \bar{K}^{*0} & \phi
\end{array}\right)_\mu\,,\label{eq:V}
\end{equation}
the $V\Phi\gamma$ interaction Lagrangian is given by~\cite{Terschluesen:2010ik}
\begin{equation}
\label{eq:VPhigamma-general}
\L_{V\Phi\gamma}=\frac{ed_{V}}{4}\epsilon^{\mu\nu\alpha\beta}F_{\mu\nu}\Tr\left(\{Q,V_\alpha\}\partial_\beta\Phi\right)\,,
\end{equation}
with the quark charge matrix
$
Q=\text{diag}(2,-1,-1)/3
$
and the electromagnetic field strength tensor
$
F_{\mu\nu}=\partial_\mu A_\nu - \partial_\nu A_\mu
$.
Expanding Eq.~\eqref{eq:VPhigamma-general}, we find the interaction Lagrangian for $V\pi^0 \gamma$ vertices
\begin{align}
\L_{V\pi^0\gamma}&=\frac{e}{2}\epsilon^{\mu\nu\alpha\beta}F_{\mu\nu}\big[d_V (c_\rho \rho^0_\alpha + c_\omega \omega_\alpha) + 
\bar{d}_{\phi} \phi_\alpha\big]\partial_\beta\pi^0 \,,
\label{eq:radiative-coupling-L}
\end{align}
where the Clebsch--Gordan coefficients $c_\rho=1/3$,  $c_\omega=1$ account for $SU(3)$ symmetry, and we have added
an OZI-suppressed~\cite{Okubo:1963fa,Zweig1964,Iizuka:1966wu} $\phi\pi^0 \gamma$ coupling $\bar{d}_{\phi}$ by hand, normalized in analogy to the $\omega\pi^0\gamma$ vertex. The couplings $d_{V}$ are fixed via the corresponding partial widths
\begin{equation}
\Gamma(V\rightarrow\pi^0\gamma)=\frac{e^2c_V^2d_{V}^2}{96\pi}\left(\frac{M_V^2-M_\pi^2}{M_V}\right)^{3}\,. \label{eq:Gamma-VPg}
\end{equation}
For the results listed in Table~\ref{tab:coupling-constants}, we have determined the couplings $d_\omega$, $d_\rho^{\text{n/c}}$ separately, which however agree with each other within uncertainties.
\begin{table}
\renewcommand{\arraystretch}{1.3}
    \centering
    \begin{tabular}{lll}
    \toprule
    &Partial width & Coupling constant\\\midrule
	$\omega\rightarrow\pi^0\gamma$& $713.2(19.9)\,\text{keV}$ & $|d_{\omega}| = \SI{2.32\pm 0.04}{GeV^{-1}}$\\
	$\rho^0\rightarrow\pi^0\gamma$ & $70.1(9.0)\,\text{keV}$ & $|d^\text{n}_{\rho}| =\SI{2.22\pm 0.15}{GeV^{-1}}$\\
	$\rho^\pm\rightarrow\pi^\pm\gamma$ & $66.5(7.4)\,\text{keV}$ & $|d^\text{c}_{\rho}| =\SI{2.16\pm 0.12}{GeV^{-1}}$\\
	$\phi\rightarrow\pi^0\gamma$ & $\SI{5.5\pm0.2}{keV}$ & $|\bar{d}_{\phi}|=\SI{0.135\pm 0.003}{GeV^{-1}}$\\
	$\phi\rightarrow K^+K^-$ & $\SI{2.091\pm.022}{MeV}$ & $|g^\text{c}_\phi| = \SI{6.33\pm0.03}{} $\\
	$\phi\rightarrow K^0_LK^0_S $ & $\SI{1.445\pm.018}{MeV}$ & $|g^\text{n}_\phi| = \SI{6.48\pm0.04}{} $\\
	$\omega\to K\bar K$&& $|g_\omega| = \SI{7.1\pm 0.8}{} $\\
	$a_2\rightarrow K\bar{K}$ & $\SI{5.2\pm 0.9}{MeV}$ & $|g_T| = 22(2)\MeV$\\
	$a_2\rightarrow \gamma\pi$ & $\SI{311\pm 032}{keV}$ & $|c_T| = \SI{0.062\pm 0.003}{GeV^{-1}}$ \\\bottomrule
    \end{tabular}
    \caption{The absolute values of the coupling constants required for the tree-level resonance exchange calculations are determined via the partial widths~\cite{Zyla:2020zbs} of the corresponding two-body decays.
    }
    \label{tab:coupling-constants}
\renewcommand{\arraystretch}{1.0}
\end{table}

The interaction Lagrangian for $V\Phi\Phi$ vertices reads
\begin{equation}
\label{eq:VPP-Lagrangian}
\L_{V\Phi\Phi}=\frac{g_V}{4}\Tr\left(V^\mu\left[\partial_\mu\Phi,\Phi\right]\right).
\end{equation}
We fix the coupling $g^\text{c}_\phi$ via the $\phi\rightarrow K^+K^-$ partial width 
\begin{equation}
\Gamma\left(\phi\rightarrow K^+K^-\right)=\frac{(g^\text{c}_\phi)^2}{96\pi}M_\phi\left(1-\frac{4M_{K^\pm}^2}{M_\phi^2}\right)^{\frac{3}{2}}\,. \label{eq:Gamma-phi-KK}
\end{equation}
The result is also shown in Table~\ref{tab:coupling-constants}, and deviates slightly from the one for neutral kaons. Both agree with the $\rho\pi\pi$ coupling at the 5\% level, $|g_\rho| = 6.01\big({}^{+0.04}_{-0.07}\big)$~\cite{GarciaMartin:2011jx}.
Note that we neglect complex phases for the $t$-channel couplings throughout.

The $\omega\to K\bar{K}$ coupling required for the dominant $\omega$-exchange contribution cannot be determined from a direct decay. 
One option is to simply fix it using $SU(3)$ symmetry, with plausible guesses at best of the uncertainty attached.
A somewhat more data-driven access to this coupling can be obtained by relying on a vector-meson-dominance (VMD) model fitted to time-like kaon form factor data from $e^+e^-\to K^+K^-$, $K_SK_L$, and $\tau^-\to K^-K_S\nu_\tau$, see Model II in Ref.~\cite{Beloborodov:2019fmw}. Together with the $\omega$--photon coupling from $\omega\to e^+e^-$~\cite{Hoferichter:2017ftn} we obtain $g_{\omega}$ as shown in Table~\ref{tab:coupling-constants}. The error is dominated by the fit value from Ref.~\cite{Beloborodov:2019fmw}; within uncertainties, $g_{\omega}$ is indeed compatible with $SU(3)$ symmetry.

Adding the $\omega$ and $\phi$ tree-level contributions, we obtain the SVA
\begin{equation}
\label{eq:SVA-G+}
\G^{(+)}(t)=e 
\bigg[
\frac{g_\omega d_{\omega}}{M_\omega^2-t}
- \frac{\sqrt{2} g^\text{c}_\phi \bar d_{\phi}}{M_\phi^2-t}
\bigg]\,.
\end{equation}
Note that the sign difference arises due to the Clebsch--Gordan coefficents in the $V\Phi\Phi$ Lagrangian.  We use zero-width propagators for the vector mesons as $t$ is negative in $\gamma K\rightarrow K\pi$.
\end{sloppypar}

\subsection{$\G^{(0)}$ with $\pi\pi$ intermediate states}
The isospin $I=1$ $\gamma\pi\rightarrow K\bar{K}$ $P$-wave $\G^{(0)}(t)$ is dominated by the rather broad $\rho(770)$. Since the $\rho$ is a $\pi\pi$ $P$-wave resonance, we can employ a more sophisticated approach than the VMD approximation and compute this SVA dispersively, taking into account intermediate $\pi\pi$ states; see Fig.~\ref{fig:t-channel-diagram}. The corresponding unitarity relation reads
\begin{equation}
\disc \G^{(0)}(t)
= -i\frac{t}{2\sqrt{2}}
\sigma^3_\pi(t) \big[g_1^1(t)\big]^*h^1_1(t)\,,
\label{eq:unitarity-rel-t-channel-rho}
\end{equation}
with the isospin $I=1$ $P$-waves $h_1^1(t)$ for $\gamma\pi\rightarrow\pi\pi$ and $g_1^1(t)$ for $\pi\pi\rightarrow K\bar{K}$ as defined in Eqs.~\eqref{eq:PWE-gamma3pi} and \eqref{eq:PWE-t-channel}, respectively.
\begin{figure*}
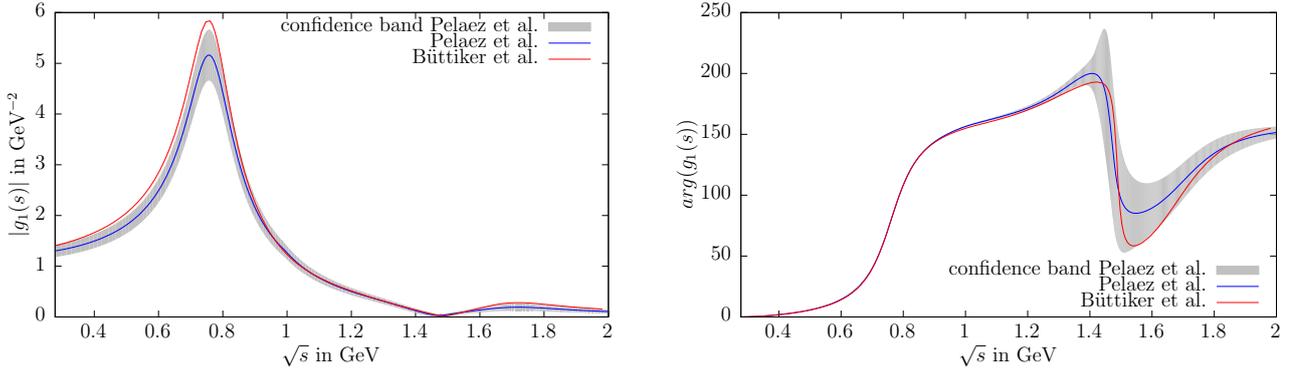

    \fontsize{12pt}{14pt} \selectfont
	\begin{subfigure}{0.5\textwidth}
		\scalebox{0.65}{\input{./plots/figures/pipiKKabs.tex}}
	\end{subfigure}
	\begin{subfigure}{0.5\textwidth}
		\scalebox{0.65}{\input{./plots/figures/pipiKKphase.tex}}
	\end{subfigure}
	\caption{The $\pi\pi\to K\bar{K}$ partial wave $g^1_1(s)$ with the parameterization from Ref.~\cite{Pelaez:2018qny} including the confidence band in blue. As a comparison, the parameterization from Ref.~\cite{Buettiker:2003pp} is shown in red.}
	\label{fig:g1compare}
\end{figure*}
\begin{figure}
    \fontsize{12pt}{14pt} \selectfont
    \centering
    \scalebox{0.65}{\input{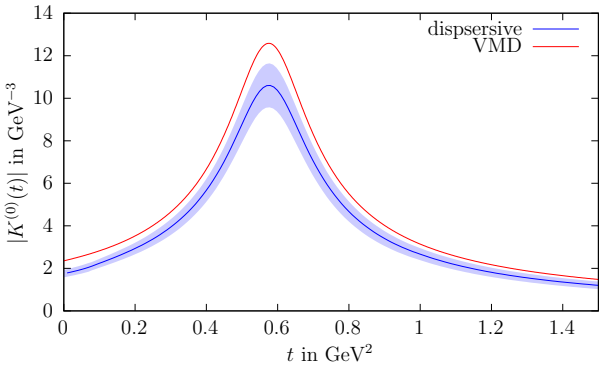}}
    \caption{Comparison of the dispersive solution and the corresponding VMD approximation for the $t$-channel SVA $\G^{(0)}$. The uncertainty band on the dispersive solution includes the propagated error of the $\pi\pi\to K\bar{K}$ amplitude $g_1^1(s)$.}
    \label{fig:rhocompare}
\end{figure}
We cast $\G^{(0)}$ into a dispersion integral 
\begin{align}
\label{eq:dispersion-integral-rho}
\G^{(0)}(t)
=
\frac{-1}{4\sqrt{2}\pi }\int_{4M_\pi^2}^\infty\dd{t'}\frac{t'\sigma^3_\pi(t')  \big[g_1^1(t')\big]^*h^1_1(t')}{t'-t} \,.
\end{align}
The input partial waves are taken from Refs.~\cite{Hoferichter:2012pm,Pelaez:2018qny} for $h_1^1(t)$ and $g_1^1(t)$, respectively.  Modulus and phase of $g_1^1(t)$ are shown in Fig.~\ref{fig:g1compare} including uncertainties, and compared to a previous parameterization~\cite{Buettiker:2003pp}.

Performing the integration is straightforward and results in the SVA $\G^{(0)}$ shown in Fig.~\ref{fig:rhocompare}. In addition, calculating the isovector $\gamma\pi\rightarrow K\bar{K}$ amplitude via neutral $\rho^0$ exchange in strict analogy to the amplitudes derived in Sec.~\ref{sec:VMD}, we obtain a VMD approximation for $\G^{(0)}$. Here, we have fixed the $\rho K\bar K$ coupling from $g_\phi^c$ and $SU(3)$ symmetry for simplicity, while the $\rho\pi\gamma$ coupling is determined via the decay $\rho^0\rightarrow\pi^0\gamma$ (which is compatible with $\rho^\pm\rightarrow\pi^\pm\gamma$ within uncertainties as required by isospin symmetry), see Table~\ref{tab:coupling-constants}.  The VMD approximation is also presented in Fig.~\ref{fig:rhocompare}. 
For the purpose of this illustration we show the amplitudes for positive $t$ and use a Breit--Wigner propagator $1/(t-M_\rho^2+iM_\rho\tilde{\Gamma}_\rho(t))$ for the simplified $\rho$ exchange. An energy-dependent width
\begin{align}
\tilde{\Gamma}_\rho(t)=\Gamma_\rho\frac{t}{M_\rho^2}\left(\frac{\sigma_\pi(t)}{\sigma_\pi(M_\rho^2)}\right)^3
\end{align}
ensures the correct threshold behavior of the phase. The VMD approximation agrees with the dispersive solution at the 20\% level in the region of the $\rho$ peak. 

\subsection{$\Ha^{(-)}$ using tensor meson dominance}

For the resonant $t$-channel $D$-wave contribution we follow the approach of Ref.~\cite{Kubis:2015sga}. We compute the SVA $\Ha^{(-)}$ via the tree-level diagram with an intermediate $a_2(1320)$ tensor meson, see Fig.~\ref{fig:t-channel-diagram}. For the interaction vertices we use the formalism presented in Ref.~\cite{Ecker:2007us}. The coupling $g_T$ for the $a_2K\bar K$ vertex is related to the partial width via
\begin{align}
\label{eq:Gamma-a2KK}
\Gamma(a_2\rightarrow K\bar{K})=
\frac{g_T^2 m_{a_2}^3}{120\pi F_\pi^4} \sigma_K^5\big(m_{a_2}^2\big)\,.
\end{align}
The extracted value $g_T=22(2)\MeV$ is about 20\% smaller than the ones extracted from $a_2\to\pi\eta$~\cite{Kubis:2015sga} or $f_2\to\pi\pi$~\cite{Ecker:2007us}, indicating a certain amount of $SU(3)$ breaking.  
Similarly, the coupling $c_T$ for the $a_2\gamma\pi$ vertex can be determined via the partial width of radiative $a_2$ decays~\cite{Zyla:2020zbs} (updated compared to Ref.~\cite{Kubis:2015sga} due to the inclusion of the new COMPASS measurement~\cite{Adolph:2014mup}),
\begin{equation}
\label{eq:Gamma-a2gammapi}
\Gamma\left(a_2\rightarrow \gamma \pi\right) = \frac{e^2c_T^2}{160\pi F_\pi^2}\frac{\left(m_{a_2}^2-M_\pi^2\right)^5}{m_{a_2}^5}\,.
\end{equation}
The results are shown in Table~\ref{tab:coupling-constants}. The tree-level contribution of the $a_2$ meson reads
\begin{equation}
\label{eq:SVA-H-}
\Ha^{(-)}(t)=\frac{2\sqrt{2}ec_Tg_T}{F_\pi^3}\frac{1}{m_{a_2}^2-t}\,.
\end{equation}
Following the arguments presented in Ref.~\cite{Kubis:2015sga} we fix the signs of the coupling constants via 
\begin{equation}
c_Tg_T=+|c_Tg_T|.
\end{equation}
The high-energy behavior of the $t$-channel amplitudes will become important in the following sections. We observe that the $P$-waves run like $1/t$ and the $D$-wave behaves like a constant due to the kinematic factor in front of $\Ha^{(-)}(t)$.

\section{Dispersive representations and Khuri--Treiman solutions}\label{sec:disp}

In this section, we discuss the main part of the dispersive representation of the $\gamma K\to K\pi$ amplitudes, the reconstruction of the $s$- and $u$-channel partial waves or SVAs.
This consistently incorporates $K\pi$ $P$-wave rescattering in the elastic approximation.

From the reconstruction theorems in Sec.~\ref{sec:isospin}, we can obtain the relevant partial waves, i.e., the $P$-waves of different isospins, via Eq.~\eqref{eq:Pwave-projection} with the result
\begin{equation}
\label{eq:decomposition-inhomo}
f^{(i)}_1(s)=\F^{(i)}(s)+\widehat{\F}^{(i)}(s)\,, \quad
i=0,\,1/2,\,3/2\,.
\end{equation}
Here, the functions $\widehat{\F}^{(i)}(s)$ originate from the $t$- and $u$- channel SVAs; they only contribute left-hand cuts to the partial waves and hence have no discontinuities along the right-hand cut. 
The $\widehat{\F}^{(i)}(s)$ can be collected from appropriate linear combinations of the different reconstruction theorems, resulting in
\begin{align}
\widehat{\F}^{(0)}(s)&=\frac{3}{4}\int_{-1}^{1}\dd{z_s}\left(1-z_s^2\right)\left[\G^{(0)}(t)+\F^{(0)}(u)\right]\,, \notag
\\
\widehat{\F}^{(1/2)}(s)&=\frac{1}{2}\int_{-1}^{1}\dd{z_s}\left(1-z_s^2\right)\bigg[\frac{1}{2}\G^{(+)}(t) \notag\\
+(u &-s)\Ha^{(-)}(t)-\frac{1}{2}\F^{(1/2)}(u)+\F^{(3/2)}(u)\bigg]
,\notag
\\
\widehat{\F}^{(3/2)}(s)&=\frac{1}{2}\int_{-1}^{1}\dd{z_s}\left(1-z_s^2\right)\bigg[\G^{(+)}(t)  \notag\\
-(u &-s)\Ha^{(-)}(t)+2\F^{(1/2)}(u)+\frac{1}{2}\F^{(3/2)}(u)\bigg]
.
\label{eq:inhomogeneities}
\end{align}
We observe that the isoscalar and isovector parts of the photon completely decouple from each other, since the isovector inhomogeneities only depend on the isovector SVAs and vice versa for the isoscalar part.

\begin{figure}[t]
	\centering
	\begin{tikzpicture}
	\begin{feynman}
	\vertex(a1) {$\gamma$};
	\vertex[below=10em of a1](a2) {$K$};
	\vertex[below=5em of a1](a3) {};
	\vertex[right=5em of a3,dot](b1) {};
	\vertex[right=10em of b1,dot](b2) {};
	\vertex[right=5em of b2](c3) {};
	\vertex[right=20em of a1](c1) {$K$};
	\vertex[below=5em of c3](c2) {$\pi$};
	\vertex[right=10em of a1](d1) {};
	\vertex[right=10em of a2](d2) {};
	\diagram* {
		(a1)--[photon](b1),(a2)--(b1),(b1)--[bend left=45, edge label=$K$, pos=0.45](b2),(b2)--[bend left=45, edge label=$\pi$, pos=0.55](b1),(b2)--(c1),(b2)--(c2),(d1)--[scalar](d2)
	};
	\end{feynman}
	\end{tikzpicture}
	\caption{Diagram of the $s$-channel contribution to $\gamma K\to K \pi$.}
	\label{fig:s-channel-diagram}
\end{figure}
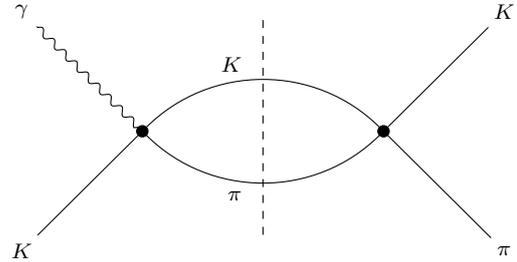

In the approximation of elastic unitarity, a right-hand cut in the amplitude is induced by intermediate $K\pi$ states, see Fig.~\ref{fig:s-channel-diagram}. 
Along the right-hand cut, the corresponding unitarity relation for the $P$-waves $f^{(i)}_1(s)$ reads
\begin{equation}
\Im f^{(i)}_1(s) = \kappa(s) \left[t^I_1(s)\right]^* f^{(i)}_1(s)\ \theta\left(s-\sth\right)\,,
\label{eq:unitarityrel}
\end{equation}
with the $K\pi$ $P$-wave amplitude $t^I_1(s)$. 
Here, the partial waves for $i=0,\,1/2$ are both associated with $I=1/2$, 
while $i=3/2$ requires $I=3/2$. 
The unitarity relation implies Watson's final-state theorem, which states that the phase of $f^{(i)}_1(s)$ coincides with $\delta_1^I(s)$~\cite{Watson:1954uc}. 
Inserting Eq.~\eqref{eq:decomposition-inhomo} and remembering that the $\widehat{\F}^{(i)}(s)$ are free of right-hand-cut discontinuities, we find a unitarity relation for the SVAs,
\begin{equation}
\Im \F^{(i)}(s) = \left(\F^{(i)}(s)+\widehat{\F}^{(i)}(s)\right)e^{-i\delta^I_1(s)}\sin\delta^I_1(s)\,.
\label{eq:ImF}
\end{equation}
Due to Eq.~\eqref{eq:ImF}, the functions $\widehat{\F}^{(i)}(s)$ are usually referred to as inhomogeneities, as they constitute the inhomogeneous contributions to the unitarity relations for $\F^{(i)}(s)$. 

\begin{figure*}
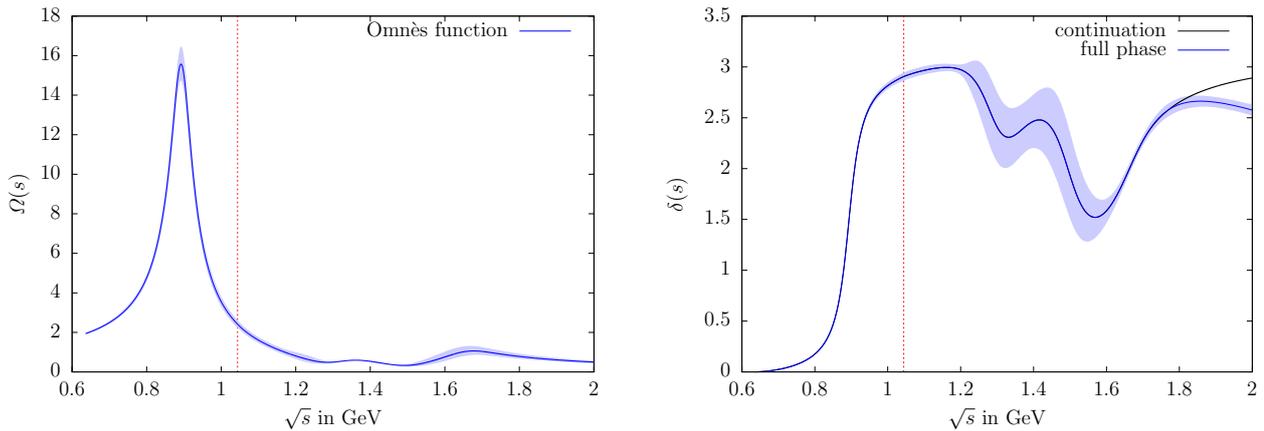

    \fontsize{12pt}{14pt} \selectfont
	\begin{subfigure}{0.49\textwidth}
		\centering
		\scalebox{0.65}{\input{./plots/figures/Omnes.tex}}
	\end{subfigure}
	\begin{subfigure}{0.49\textwidth}
		\centering
		\scalebox{0.65}{\input{./plots/figures/phase.tex}}
	\end{subfigure}
	\caption{Absolute value and phase of the \Omnes function for the isospin $I=1/2$ $P$-wave $K\pi$ scattering phase shift. The dashed red line indicates the $K\eta$ threshold. The phase incorporates the higher resonances $K^*(1410)$ and $K^*(1680)$ and is continued towards $\pi$ starting at $\sqrt{s}=1.7\GeV$ shown in the black line.}
	\label{fig:Omnes}
\end{figure*}

A solution to the corresponding homogeneous unitarity relation, setting the inhomogeneities to zero, is given by the \Omnes function
\begin{equation}
\Omega^I(s)=\exp\left(\frac{s}{\pi}\int_{\sth}^\infty\frac{\dd s'}{s'}\frac{\delta^I_1(s')}{(s'-s)}\right)\,,
\end{equation}
determined entirely by the corresponding $K\pi$ phase shifts.
The input for the latter is taken from Ref.~\cite{Pelaez:2016tgi}. We smoothly continue the $I=1/2$ phase shift to $\pi$ starting at $\sqrt{s}=\SI{1.7}{\GeV}$.  As a result, we obtain the high-energy behavior of the corresponding Omn\`es function $\Omega^{1/2}(s\to\infty) \propto 1/s$.
Since we approximate the $I=3/2$ phase shift to be zero, the \Omnes function is just $\Omega^{3/2}(s)=1$. We therefore suppress the superscript $\Omega^{1/2}(s)=\Omega(s)$ in the following. The \Omnes function and its phase are shown in Fig.~\ref{fig:Omnes}. 

The solution of the full, inhomogeneous unitarity relation~\eqref{eq:ImF} for the single-variable amplitudes is subsequently obtained using a separation ansatz with the \Omnes function. The results are the Khuri--Treiman equations~\cite{Khuri:1960zz} for the SVAs
\begin{align}
\F^{(0,1/2)}(s)&=\Omega(s)\Bigg(P^{(0,1/2)}_{n-1}(s)\notag\\
&\quad~+\frac{s^{n}}{\pi}\int_{\sth}^\infty \frac{\dd s'}{s'^{n}} \frac{\widehat{\F}^{(0,1/2)}(s')\sin\delta^{1/2}_1(s')}{|\Omega(s')|(s'-s)}\Bigg)\,,\notag\\
\F^{(3/2)}(s)&=P^{(3/2)}_{n'-1}(s)\,.
\label{eq:singlevariableamplitudes}
\end{align}
We can solve for the SVAs by inserting the fixed $t$-channel contributions from Sec.~\ref{sec:tchannel} into the inhomogeneities and then solving Eq.~\eqref{eq:singlevariableamplitudes} iteratively. The system is linear in the subtraction constants, so that it is possible to construct basis functions. The calculation of the latter converges very quickly, such that they remain practically unchanged after at most five iterations. 

Demanding that the single-variable amplitudes in Eq.~\eqref{eq:singlevariableamplitudes} have the same high-energy behavior leads to constraints on the order of the subtraction polynomials of the form $n=n'+1$. We will investigate two different subtraction schemes. The first one is $n=1$, which we will call the minimal subtraction scheme. In this case the $I=3/2$ component needs to vanish and we are not able to include the $D$-wave in the $t$-channel. As a result, the subtraction polynomials are simply given by
\begin{align}
P^{(0)}_0(s)=a^{(0)}_1 \,, \qquad
P^{(1/2)}_0(s)&=a^{(1/2)}_1\,.
\end{align}
Asymptotically, the scattering amplitudes $A^{(i)}(s,t,u)$ drop with one inverse power of the Mandelstam variables, hence no polynomial ambiguity, see Eq.~\eqref{eq:ambiguity}, can be exploited.
If we subtract one more time, we obtain $n=2$ and the $I=3/2$ component is a constant, while the isospin $I=1/2$ subtraction polynomials are linear:
\begin{align}
P^{(0)}_1(s)&=a^{(0)}_2+ b^{(0)}_2 s\,,\notag\\
P^{(1/2)}_1(s)&=a^{(1/2)}_2+b^{(1/2)}_2 s\,,\notag\\
P^{(3/2)}_0(s)&=a^{(3/2)}_2\,.
\end{align}
In this scheme, the $A^{(i)}(s,t,u)$ approach a constant for high energies.
The constant ambiguity of Eq.~\eqref{eq:ambiguity} (due to the parameter $\beta$) allows us to reduce the number of free parameters by one, as we can eliminate $a^{(3/2)}_2$ and absorb its effects into $a^{(1/2)}_2$. 
This means that we can drop the SVA $\F^{(3/2)}(s)$ also in the $n=2$ subtraction scheme, due to Eq.~\eqref{eq:singlevariableamplitudes}.

While the angular projection integrals for the inhomogeneities in Eq.~\eqref{eq:inhomogeneities} are straightforward to calculate for physical values of $s$, the continuation into the complex $s$-plane and onto the second Riemann sheet for the investigation of resonance poles and their residues is far less obvious.
For this purpose, we use an alternative kernel method~\cite{Hoferichter:2017ftn}:
we replace the SVAs by their dispersive representations, given through integrals along the right-hand discontinuities,  
and evaluate the angular integrals explicitly. 
In the twice-subtracted scheme, and neglecting for simplicity the $t$-channel $D$-wave contributions, this procedure results in
{\allowdisplaybreaks[1]
\begin{align}
f(s)&=\frac{3}{4}\int_{-1}^{1}\dd z_s\left(1-z_s^2\right)\left[A(s)+B(t)+C(u)\right] \notag\\
&=
\frac{3}{4}\int_{-1}^{1}\dd z_s\left(1-z_s^2\right)
\bigg[
c_{s,2}^{(1)} + c_{s,2}^{(2)}s \notag\\
&\qquad\qquad+\frac{1}{\pi}\int_{\sth}^{\infty}\frac{\dd s'}{s'^2}\frac{s^2}{s'-s}\Im A(s')
\bigg] \notag\\
&\phantom{=}
+
\frac{3}{4}\int_{-1}^{1}\dd z_s\left(1-z_s^2\right)
\bigg[
c_{t,2}^{(1)} + c_{t,2}^{(2)}t \notag\\
&\qquad\qquad+\frac{1}{\pi}\int_{\sth}^{\infty}\frac{\dd s'}{s'^2}\frac{t^2}{s'-t}\Im B(s')
\bigg] \notag\\
&\phantom{=}
+
\frac{3}{4}\int_{-1}^{1}\dd z_s\left(1-z_s^2\right)
\bigg[
c_{u,2}^{(1)} + c_{u,2}^{(2)}u \notag\\
&\qquad\qquad+\frac{1}{\pi}\int_{\sth}^{\infty}\frac{\dd s'}{s'^2}\frac{u^2}{s'-u}\Im C(s')
\bigg] \notag\\
&=
c_{s,2}^{(1)} + c_{s,2}^{(2)}s+\frac{1}{\pi}\int_{\sth}^{\infty}\dd s' K_2^s(s,s')\Im A(s') \notag\\
&\phantom{=}
+c_{t,2}^{(1)} + c_{t,2}^{(2)}a^{+\Delta} +\frac{1}{\pi}\int_{\sth}^{\infty}\dd s' K_2^t(s,s')\Im B(s') \notag\\
&\phantom{=}
+c_{u,2}^{(1)} + c_{u,2}^{(2)}a^{-\Delta}+\frac{1}{\pi}\int_{\sth}^{\infty}\dd s' K_2^u(s,s')\Im C(s')
\,\label{eq:kernelfunc}
\end{align}}%
for a generic partial wave $f$.  In the concrete applications, the complete reconstruction theorems from Sec.~\ref{sec:isospin} need to be inserted in place of the functions $A(s)$, $B(t)$, and $C(u)$. 
The explicit forms of the kernel functions, as well as
analogously defined kernels for the once-subtracted form and the generalization including the effect of the $t$-channel $D$-wave SVA, are given in~\ref{app:kernel}.

Using this method it is possible to calculate the partial waves in the full complex plane with the correct analytic continuation. The solution depends on two or four free parameters in the different subtraction schemes.

\section{Matching}\label{sec:matching}
In this section, we discuss how to fix the free parameters of the dispersive representation, the subtraction constants, by matching them to the chiral anomaly on the one hand, and the radiative couplings of the $K^*(892)$ resonances on the other. 

\subsection{Chiral anomaly}
\begin{sloppypar}
The Wess--Zumino--Witten anomaly~\cite{Wess:1971yu,Witten:1983tw} yields low-energy theorems for the different $\gamma K \to K \pi$ amplitudes in the limit of vanishing energies ($s=t=u=0$) and vanishing (light as well as strange) quark masses.  It contributes to the neutral-pion-production amplitudes, but not to the charge-exchange processes:
\begin{equation}
\F^{-0/00}(0,0,0)=F_{KK\pi}\,, \quad 
\F^{0-/-+}(0,0,0)=0\,,
\end{equation}
where~\cite{Ebertshaeuser2001,Hacker2008}
\begin{equation}
\label{eq:anomaly}
F_{KK\pi}=\frac{e}{4\pi^2\Fpi^3}=9.8\GeV^{-3}
\end{equation}
is given in terms of the pion decay constant $\Fpi=\SI{92.28\pm 0.03}{MeV}$ and the electric charge $e$, and is
actually identical to the similarly defined anomaly $F_{3\pi}$ for $\gamma\pi\to\pi\pi$~\cite{Adler:1971nq,Terentev:1971cso,Aviv:1971hq}.
Equation~\eqref{eq:anomaly} is quoted for fixed number of colors $N_c=3$ only; compare Ref.~\cite{Bar:2001qk} for an extended discussion on how to properly generalize this to $N_c\neq3$.
\end{sloppypar}

In contrast to $F_{3\pi}$, however, we cannot safely assume higher-order corrections to be small.  For the former, quark-mass renormalization effects were estimated to enhance $F_{3\pi}$ by $6.6(1.0)\%$~\cite{Bijnens:1989ff,Hoferichter:2012pm}, 
based on resonance saturation.  As this correction term in fact scales with $s+t+u=3M_\pi^2$, it is not unreasonable to caution that a similar effect might potentially be much larger for $\gamma K\to K\pi$.
In addition, we would naturally expect the denominator of Eq.~\eqref{eq:anomaly} to be modified according to $F_\pi^3\to F_\pi F_K^2$ at higher orders, 
which, as $F_K$ is much further away from the common chiral-limit value,
$F_K/F_\pi = \SI{1.193\pm 0.002}{}$~\cite{Zyla:2020zbs}, would be another large effect (although in the opposite direction).
Since it is hard to estimate the correlations between all these higher-order corrections, we simply estimate a resulting uncertainty of $25\%$ on $F_{KK\pi}$.  A complete next-to-leading-order calculation of all $\gamma K\to K\pi$ channels in chiral perturbation theory would certainly be highly desirable (see Refs.~\cite{Ebertshaeuser2001,Hacker2008} for partial results).

\begin{sloppypar}
Obviously, also the vanishing charge-exchange amplitude will be modified due to higher-order corrections. Since a relative error estimate is not meaningful here, we use the absolute uncertainty given for the anomaly also for the charge-exchange amplitudes.
Our combined assumption on the different amplitude normalizations in the soft-meson limits is therefore
\begin{align}
\F^{-0/00}(0,0,0)&=9.8(2.4)\GeV^{-3}\,, \notag\\
\F^{0-/-+}(0,0,0)&=0.0(2.4)\GeV^{-3}\,. \label{eq:NLETs}
\end{align}
Inserting the reconstruction theorems, we find the two subtraction constants given explicitly by
\begin{align}
a_n^{(0)}&=\frac{1}{2}\left[\F^{0-/-+}(0,0,0)-\G^{(0)}(0)\right] \notag\\
& =0.9(1.2)\GeV^{-3}\,, \notag\\
a_n^{(1/2)}&=\frac{1}{2}\left[\F^{-0/00}(0,0,0)-\G^{(+)}(0)\right] \notag\\
& =1.0(1.3)\GeV^{-3}\,, \qquad n\in\{1,2\}\,.
\label{eq:anomaly-matching}
\end{align}
Here we have used $\G^{(+)}(0) = e\big(g_\omega d_{\omega}/M_\omega^2
- \sqrt{2} g_\phi d_{\phi}/M_\phi^2\big) =  7.8(8) \GeV^{-3}$, cf.\ Eq.~\eqref{eq:SVA-G+}, where the 
error is entirely dominated by $g_\omega$, see Table~\ref{tab:coupling-constants}.
The $\rho$ contribution $\G^{(0)}(0)$ can be evaluated by a sum rule based on Eq.~\eqref{eq:dispersion-integral-rho},
\begin{align}
\G^{(0)}(0)
&=
\frac{-1}{4\sqrt{2}\pi }\int_{4M_\pi^2}^\infty\dd{t'}\sigma_\pi(t')^3 \big[g_1^1(t')\big]^*h^1_1(t') \notag\\
&= -1.7(2)\GeV^{-3}\,.
\end{align}
The uncertainty is given by the one in the $\pi\pi\to K \bar{K}$ $P$-wave~\cite{Pelaez:2018qny}, which at the same time covers the difference to the alternative parametrization of Ref.~\cite{Buettiker:2003pp}, see Fig.~\ref{fig:g1compare}.
\end{sloppypar}

\subsection{Radiative couplings of the $K^*(892)$}
In the narrow-width approximation, the radiative widths of the $K^*(892)$ vector mesons are given by
\begin{align}
\begin{split}
\frac{1}{4}\Gamma_{K^{*0}\rightarrow K^0\gamma}=
\Gamma_{K^{*\pm}\rightarrow K^\pm\gamma}&=\frac{e^2d_{K^*}^2}{864\pi }\left(\frac{M_{K^*}^2-M_K^2}{M_{K^*}}\right)^3\,,
\end{split}
\end{align}
where the conventions for the coupling constants $d_{K^*}$ are chosen such that they coincide with $d_\omega$ in the $SU(3)$ symmetry limit, see Eq.~\eqref{eq:Gamma-VPg} and Table~\ref{tab:coupling-constants}.
The Particle Data Group~\cite{Zyla:2020zbs} lists only three measurements from which these radiative widths have been extracted, one for $K^{*0}\to K^0\gamma$~\cite{Carlsmith:1985ep} and two for $K^{*\pm}\to K^\pm\gamma$~\cite{Chandlee:1983hf,Berg1983}.
The extracted charged and neutral radiative couplings read 
\begin{equation}
\gc = \SI{2.50 \pm 0.12}{GeV^{-1}}\,, \quad 
\gn = \SI{1.93 \pm 0.08}{GeV^{-1}}\,,
\label{eq:g-charged-neutral}
\end{equation}
and thus violate $SU(3)$ symmetry at the 20\% level.

For the further discussion, we require amplitudes, and hence radiative couplings, of definite (photon) isospin, $d_{K^*}^{(0)}$ and $d_{K^*}^{(1/2)}$. 
To this end, we use the decomposition of the quark charge matrix into its isoscalar and isovector components via the Gell-Mann matrices according to
\begin{align}
Q=\frac{1}{2}\lambda_3+\frac{1}{2\sqrt{3}}\lambda_8\,,
\end{align}
which leads to the relations
\begin{align}
d^{(0)}_{K^*}&=2\gn-\gc=\SI{1.36\pm 0.2}{GeV^{-1}}\,,\notag\\
d^{(1/2)}_{K^*}&=\frac{1}{3}\left(2\gn+\gc\right)=\SI{2.12\pm 0.06}{GeV^{-1}}\,. \label{eq:g-isospin}
\end{align}
In the $SU(3)$ symmetry limit, all couplings would be equal: $d_{K^*}^{(0)}=d_{K^*}^{(1/2)}=\gc=\gn$.  Phenomenologically, Eq.~\eqref{eq:g-isospin} even enhances the symmetry breaking found in the physical charge channels, Eq.~\eqref{eq:g-charged-neutral}.
As our formalism only relies on isospin symmetry and not on $SU(3)$, there is no a priori problem in incorporating such seemingly large $SU(3)$ violation.

\begin{sloppypar}
For a model-independent extraction of the radiative $K^*$ coupling constants, we have to analytically continue the $\gamma K\to K\pi$ amplitudes onto the second Riemann sheet and connect them to the residues of the corresponding poles.   
The continuation to the second sheet can be found from the discontinuity
in Eq.~\eqref{eq:unitarityrel},
\begin{equation}
\label{eq:unitarityrel-cont}
f^{(i)}_{1,I}(s)-f^{(i)}_{1,II}(s)=-2\hat{\kappa}(s)t^{1/2}_{1,II}(s)f^{(i)}_{1,I}(s)\,,
\end{equation}
where $I$ $(II)$ denotes the first (second) Riemann sheet, and
\begin{equation}
\hat{\kappa}(s)=\frac{\sqrt{-\lambda(s,M_\pi^2,M_K^2)}}{s}\,, \quad
\hat{\kappa}(s\pm i\epsilon)=\mp i\kappa(s)\,,
\end{equation}
which leads to the correct analytic structure including the branch cut.
In the vicinity of the pole we have~\cite{Stamen2020}
\begin{align}
t^{1/2}_{1,II}(s) &= \frac{g_{K^*}^2 s\,\kappa^2(s)}{64\pi(s_{K^*}-s)}\,,\notag\\
f^{(0)}_{1,II}(s) &= \frac{1}{3}\frac{e d_{K^*}^{(0)}g_{K^*}}{s_{K^*}-s}\,,\notag\\
f^{(1/2)}_{1,II}(s)& = \frac{e d_{K^*}^{(1/2)}g_{K^*}}{s_{K^*}-s}\,,
\end{align}
with conventions for the coupling constants chosen such that they match onto the VMD expressions with the Lagrangians introduced in Eqs.~\eqref{eq:VPhigamma-general} and \eqref{eq:VPP-Lagrangian}. 
Inserting these in Eq.~\eqref{eq:unitarityrel-cont} we find 
\begin{align}
\frac{f^{(i)}_{1,II}(\sKs)}{t^{1/2}_{1,II}(\sKs)}
&=\frac{ec_{K^*}^{(i)} d_{K^*}^{(i)}}{g_{K^*}}  \frac{128\pi}{\sKs\kappa^2(\sKs)} \notag\\
&=2i\kappa(\sKs)f^{(i)}_{1,I}(\sKs)\,, \label{eq:couplingratio}
\end{align}
where $c_{K^*}^{(0)} = 1/6$ and $c_{K^*}^{(1/2)} = 1/2$ take the two different normalizations into account.
This is the connection between the (radiative) couplings and the dispersive solution, depending linearly on the subtraction constants and evaluated on the physical (first) Riemann sheet only.  
Taking the $K^*\to K\pi$ coupling $|g_{K^*}|=\SI{6.07\pm 0.67}{}$ with a small phase of $\text{arg}(g_{K^*})=\SI{-3.1\pm0.2}{\degree}$ and the $K^*$ pole position $s_{K^*}=\SI{0.7948\pm 0.002}{}-i\ \SI{0.0515\pm 0.002}{\GeV\squared}$ from a dispersive analysis of $K\pi$ scattering~\cite{Pelaez:2016klv,Elvira2019}, all other parameters are determined. With the basis functions and the kernel method from Sec.~\ref{sec:disp} we are able to calculate the partial waves at the pole position as linear functions of the subtraction constants. The errors are the propagated uncertainties of the pole position not including other sources, as for example the phase shift or $t$-channel parameters. 
For the minimal subtraction scheme we obtain
\begin{align}
f^{(0)}(s_{K^*})&=a_1^{(0)}\left[\SI{1.58\pm 0.17}{}-\SI{8.59\pm 0.16}{}i\right]\notag\\
&\quad-\left[\SI{0.914\pm 0.02}{}-\SI{1.046\pm 0.02}{}i\right]\text{GeV}^{-3}\,,\notag\\
f^{(1/2)}(s_{K^*})&=a_1^{(1/2)}\left[\SI{0.88\pm 0.15}{}-\SI{7.16\pm 0.14}{}i\right]\notag\\
&\quad+\left[\SI{1.427\pm 0.022}{}-\SI{1.767\pm 0.021}{}i\right]\text{GeV}^{-3}\,, \label{eq:fatpolemin}
\end{align}
while the twice subtracted case yields
\begin{align}
f^{(0)}(s_{K^*})&=
a_2^{(0)}\left[\SI{1.53\pm 0.17}{}-\SI{7.94\pm 0.16}{}i\right]\notag\\
&\quad+b_2^{(0)}\left[\SI{0.39\pm 0.11}{}-\SI{5.71\pm 0.11}{}i\right]
\text{GeV}^{2}\notag\\
&\quad-\left[\SI{0.869\pm 0.02}{}+\SI{0.383\pm 0.02}{}i\right]
\text{GeV}^{-3}\,,\notag\\
f^{(1/2)}(s_{K^*})&
=a_2^{(1/2)}\left[\SI{0.89\pm 0.15}{}-\SI{7.41\pm 0.14}{}i\right]\notag\\
&\quad+b_2^{(1/2)}\left[\SI{0.47\pm 0.12}{}-\SI{6.16\pm 0.12}{}i\right]
\text{GeV}^{2}\notag\\
&\quad+\left[\SI{1.336\pm 0.022}{}-\SI{0.580\pm 0.021}{}i\right]
\text{GeV}^{-3}\,, \label{eq:fatpoletwice}
\end{align}
omitting the $a_2$ contribution. Note that the subtraction-constant-independent terms originate from the $t$-channel contributions. Including the $a_2$ resonance changes the isovector-photon amplitude only,
\begin{align}
f^{(1/2)}(s_{K^*})&
=a_2^{(1/2)}\left[\SI{0.89\pm 0.15}{}-\SI{7.41\pm 0.14}{}i\right]\notag\\
&\quad+b_2^{(1/2)}\left[\SI{0.47\pm 0.12}{}-\SI{6.16\pm 0.12}{}i\right]
\text{GeV}^{2} \notag\\
&\quad+\left[\SI{1.091\pm 0.012}{}+\SI{0.451\pm 0.012}{}i\right]
\text{GeV}^{-3}\,. 
\end{align}
We do not quote the uncertainties due to the $t$-channel contributions separately as, in any fit to experimental data, these will be strongly correlated with the subtraction constants; see, e.g., Eq.~\eqref{eq:anomaly-matching}. 
\end{sloppypar}

\section{Results}\label{sec:results}

We begin the discussion of numerical results with the minimal subtraction scheme, which contains two subtractions constants.  According to the discussion of the previous section, we can choose to fix these in two different ways: via matching to the chiral anomaly or, more precisely, our estimate of the amplitudes at the kinematical point $s=t=u=0$, cf.\ Eq.~\eqref{eq:anomaly-matching}; 
or by reproducing the experimentally measured radiative $K^*$ couplings via Eqs.~\eqref{eq:couplingratio} and \eqref{eq:fatpolemin}.

\begin{figure*}
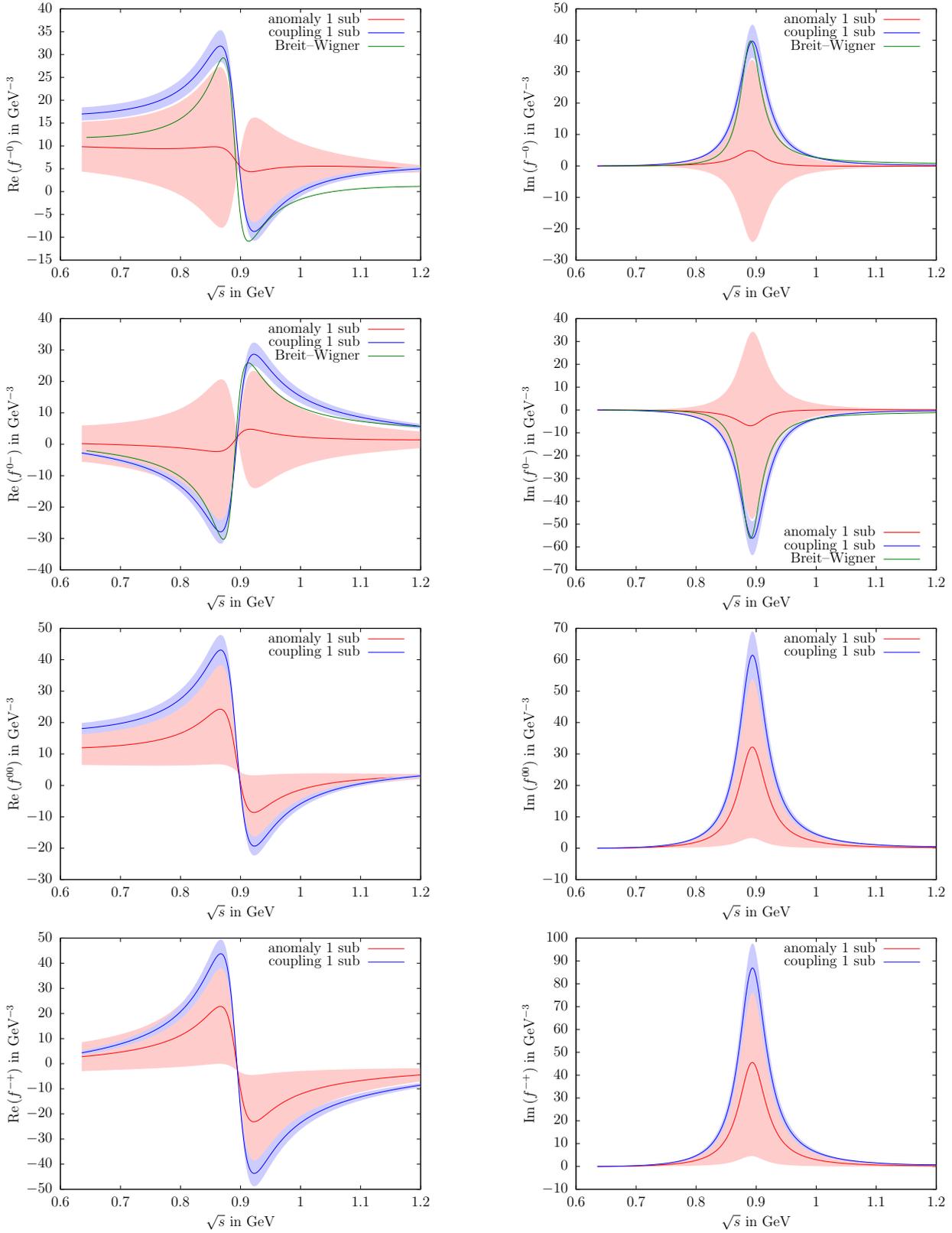

    \fontsize{12pt}{14pt} \selectfont
	\begin{subfigure}{0.49\textwidth}
	    \scalebox{0.6}{\input{./plots/amplitudes/F1_1sub_Re.tex}}
	\end{subfigure}
	\hfill
	\begin{subfigure}{0.49\textwidth}
		\scalebox{0.6}{\input{./plots/amplitudes/F1_1sub_Im.tex}}
	\end{subfigure}
	\hfill
	\begin{subfigure}{0.49\textwidth}
		\scalebox{0.6}{\input{./plots/amplitudes/F2_1sub_Re.tex}}
	\end{subfigure}
	\hfill
	\begin{subfigure}{0.49\textwidth}
		\scalebox{0.6}{\input{./plots/amplitudes/F2_1sub_Im.tex}}
	\end{subfigure}
	\hfill
	\begin{subfigure}{0.49\textwidth}
		\scalebox{0.6}{\input{./plots/amplitudes/F3_1sub_Re.tex}}
	\end{subfigure}
	\hfill
	\begin{subfigure}{0.49\textwidth}
		\scalebox{0.6}{\input{./plots/amplitudes/F3_1sub_Im.tex}}
	\end{subfigure}
	\hfill
	\begin{subfigure}{0.49\textwidth}
		\scalebox{0.6}{\input{./plots/amplitudes/F4_1sub_Re.tex}}
	\end{subfigure}
	\hfill
	\begin{subfigure}{0.49\textwidth}
		\scalebox{0.6}{\input{./plots/amplitudes/F4_1sub_Im.tex}}
	\end{subfigure}
	\caption{Results for $\gamma K^-\to K^- \pi^0$, $\gamma K^-\to \bar{K}^0 \pi^-$, $\gamma \bar{K}^0\to \bar{K}^0 \pi^0$, and $\gamma \bar{K}^0\to K^- \pi^+$ $P$-wave amplitudes (from top to bottom) in the minimal subtraction scheme, matched to the chiral anomaly and the $K^*$ radiative couplings separately. Left and right panels show real and imaginary parts of the partial waves, respectively. For comparison a Breit--Wigner model~\cite{Vysotsky:2015mks} is also shown. The error bands refer to the uncertainties on the subtraction constants and the $t$-channel amplitudes.}
	\label{fig:f_1sub}
\end{figure*}

We start with the first option and match the subtraction constants to the low-energy theorems.  Figure~\ref{fig:f_1sub} shows real and imaginary parts of the full $s$-channel partial waves for all four charge configurations, and compares these to a tree-level model for the two $\gamma K^-$ channels with Breit--Wigner propagators for all vector exchanges taken from Ref.~\cite{Vysotsky:2015mks}.  We propagate the uncertainties both on the subtraction constants and the $t$-channel amplitudes. 
Obviously, the error bands are huge; a strong deviation of the central values from the tree-level model is also observed.  This illustrates the very strong dependence of the partial waves, and in particular the $K^*(892)$ resonance signals, on the amplitudes in the low-energy limit.  By reversing the argument, a concise measurement of the cross section around the resonance peak will help determine the anomaly and, potentially, its higher-order corrections very accurately if the minimal subtraction scheme can be validated experimentally to be sufficient.  This is in strict analogy to the argument of Ref.~\cite{Hoferichter:2012pm} that the full resonance signal of the $\rho(770)$ can be employed to extract the chiral anomaly in $\gamma\pi\to\pi\pi$.

We note that the anomaly-matched amplitude for $\gamma K^- \to K^- \pi^0$ shows no zero crossing in the real part around the $K^*(892)$, since it includes the $I=3/2$ partial wave as well.  The latter is non-vanishing due to its inhomogeneity also for a vanishing $I=3/2$ single-variable amplitude; cf.\  Eq.~\eqref{eq:inhomogeneities}. Watson's final-state theorem is fulfilled for the $I=1/2$ amplitudes by construction. This effect is incorporated in all partial waves for physical charge channels with the different subtraction schemes, but only easily discernible in Fig.~\ref{fig:f_1sub}. 
The radiative $K^*$ couplings calculated from this solution correspondingly have huge uncertainties: 
\begin{align}
d_{K^*}^{(0)}&=\left[1.7(2.6)+0.06(14)i\right]\GeV^{-1}\,,\notag\\
d_{K^*}^{(1/2)}&=\left[0.7(7)-0.2(6)i\right]\GeV^{-1}\,.
\end{align}
The central value for the isoscalar coupling is closer to the experimental result, while the isovector one differs from its phenomenological value by two standard deviations. The large errors reflect the variation of the $\omega$-exchange amplitude, which is used explicitly in the matching, and the higher-order corrections to the anomaly, which we have estimated rather conservatively.
Furthermore the complex phases are small as expected (cf.\ analogous $\rho$ couplings \cite{Hoferichter:2017ftn}), but the signs cannot be determined due to the large uncertainties.
Overall, the tendency towards smaller radiative couplings is in agreement with Fig.~\ref{fig:f_1sub}, where the $K^*$ resonance peaks are seen to be significantly less pronounced than what the tree-level model predicts in the $\gamma K^-$ amplitudes~\cite{Vysotsky:2015mks}.

As the second approach, we fix the real subtraction constants in the minimal subtraction scheme using the radiative $K^*$ couplings derived from experiment.  This results in
\begin{align}
a_1^{(0)}= \SI{0.74\pm 0.12}{\per\GeV\cubed}\,,\qquad
a_1^{(1/2)}=\SI{3.18\pm 0.42}{\per\GeV\cubed}\,.
\end{align}
Real and imaginary parts of the resulting partial waves are also shown in Fig.~\ref{fig:f_1sub} for all four charge configurations. We observe that the uncertainties are much smaller in this scheme. The heights of the resonance peaks match the simple Breit--Wigner parameterization~\cite{Vysotsky:2015mks} around the resonance mass.
When calculating the different amplitudes at $s=t=u=0$ for these subtraction constants, we find
\begin{align}
\F^{-0}(0,0,0)&=14.4(1.8)\GeV^{-3}\,,\notag\\
\F^{0-}(0,0,0)&=\SI{-1.4 \pm 0.4}{\per\giga\electronvolt\cubed}\,,\notag\\
\F^{00}(0,0,0)&=13.8(2.0)\GeV^{-3}\,,\notag\\
\F^{-+}(0,0,0)&=\SI{-1.4 \pm 0.4}{\per\giga\electronvolt\cubed}\,.
\end{align}
While the low-energy limits for the charge-exchange amplitudes, to which the Wess--Zumino--Witten anomaly does not contribute, come out well within the range we have estimated roughly due to possible higher-order corrections, see Eq.~\eqref{eq:NLETs}, the amplitudes for $\pi^0$ production exceed the value of the chiral anomaly by 40\%--50\%, or $2$--$2.5\sigma$ in terms of the experimental uncertainties.  This confirms the statement we have made above: if the minimal subtraction scheme turns out to be sufficient to describe the cross section data up to and around the $K^*$ resonance, even the rather time-honored data of Refs.~\cite{Berg1983,Chandlee:1983hf,Carlsmith:1985ep} will allow us to draw strong conclusions on quark mass corrections to the chiral amplitudes for $\gamma K\to K\pi$.

\begin{figure*}
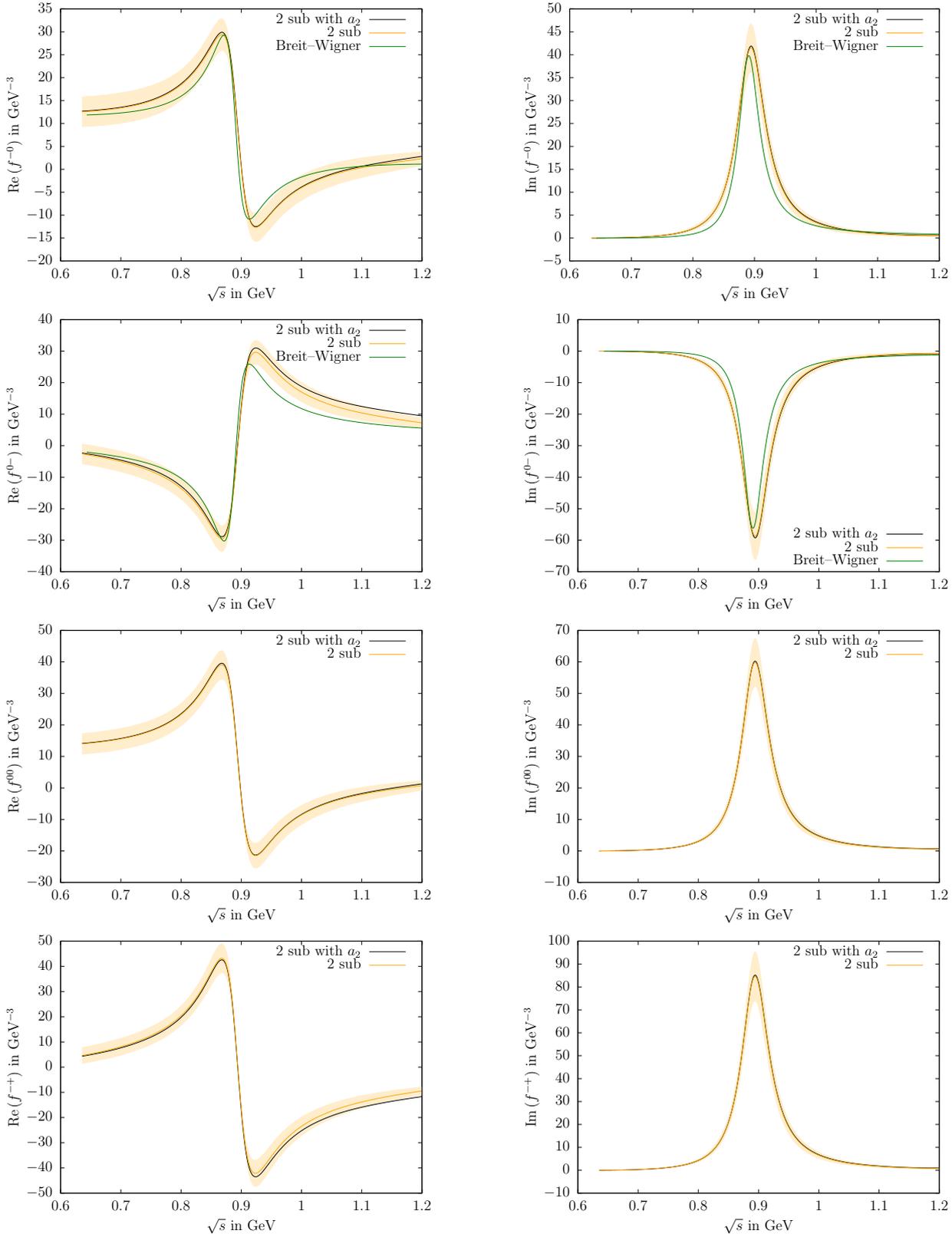

    \fontsize{12pt}{14pt} \selectfont
	\begin{subfigure}{0.49\textwidth}
		\scalebox{0.6}{\input{./plots/amplitudes/F1_2sub_Re.tex}}
	\end{subfigure}
	\hfill
	\begin{subfigure}{0.49\textwidth}
	    \scalebox{0.6}{\input{./plots/amplitudes/F1_2sub_Im.tex}}
	\end{subfigure}
	\hfill
	\begin{subfigure}{0.49\textwidth}
		\scalebox{0.6}{\input{./plots/amplitudes/F2_2sub_Re.tex}}
	\end{subfigure}
	\hfill
	\begin{subfigure}{0.49\textwidth}
		\scalebox{0.6}{\input{./plots/amplitudes/F2_2sub_Im.tex}}
	\end{subfigure}
	\hfill
	\begin{subfigure}{0.49\textwidth}
		\scalebox{0.6}{\input{./plots/amplitudes/F3_2sub_Re.tex}}
	\end{subfigure}
	\hfill
	\begin{subfigure}{0.49\textwidth}
		\scalebox{0.6}{\input{./plots/amplitudes/F3_2sub_Im.tex}}
	\end{subfigure}
	\hfill
	\begin{subfigure}{0.49\textwidth}
		\scalebox{0.6}{\input{./plots/amplitudes/F4_2sub_Re.tex}}
	\end{subfigure}
	\hfill
	\begin{subfigure}{0.49\textwidth}
		\scalebox{0.6}{\input{./plots/amplitudes/F4_2sub_Im.tex}}
	\end{subfigure}
	\caption{Results for $\gamma K^-\to K^- \pi^0$, $\gamma K^-\to \bar{K}^0 \pi^-$, $\gamma \bar{K}^0\to \bar{K}^0 \pi^0$, and $\gamma \bar{K}^0\to K^- \pi^+$ $P$-wave amplitudes (from top to bottom), matched to the anomaly and the couplings simultaneously in the twice subtracted scheme, with and without the $a_2$ resonance. Left and right panels show real and imagnary parts of the partial waves, respectively. For comparison a Breit--Wigner model~\cite{Vysotsky:2015mks} is also shown. For the solution without the $a_2$ resonance, the error bands refer to the uncertainties on the subtraction constants and the $t$-channel amplitudes.}
	\label{fig:f_2sub}
\end{figure*}

To obtain a larger degree of flexibility for the description of future high-precision cross section data, 
we can apply the twice subtracted version with four degrees of freedom.  This allows us to include both constraints, low-energy theorems and resonance couplings, and combine them into a prediction for experiment. Using the subtraction constants $a_2^{(i)}$ found from low-energy matching, we can calculate the remaining $b_2^{(i)}$ via the resonance couplings, with the result
\begin{align}
\label{eq:b_2-noa2}
b_2^{(0)}=-0.4(1.7)\GeV^{-5}\,, \qquad 
b_2^{(1/2)}=2.7(1.7)\GeV^{-5}\,.
\end{align}
The corresponding amplitude plots are shown in Fig.~\ref{fig:f_2sub}, again for all four charge configurations. The uncertainty bands for the partial waves are smaller due to the large anti-correlation of the errors of the two subtraction constants in each isospin channel.

Furthermore, in the twice subtracted representation it is possible to include the $a_2$ $t$-channel contribution, which changes the isovector part of the photon only. The corresponding plots are also included in Fig.~\ref{fig:f_2sub}; the updated subtraction constants $b^{(i)}_2$ read
\begin{align}
b_2^{(0)}=-0.4(1.7)\GeV^{-5}\,,\qquad 
b_2^{(1/2)}=2.9(1.7)\GeV^{-5}\,.
\end{align}
Comparing the two solutions with and without $a_2$-exchange, we observe that this mechanism is fully negligible below $\SI{1}{\GeV}$ for neutral-pion production. Furthermore, the deviations in the charge-exchange channels are inside the uncertainty bands and therefore also small. We conclude that it is unnecessary to take $D$- and higher partial waves into account when considering the left-hand cuts at the current level of accuracy.

\begin{figure*}
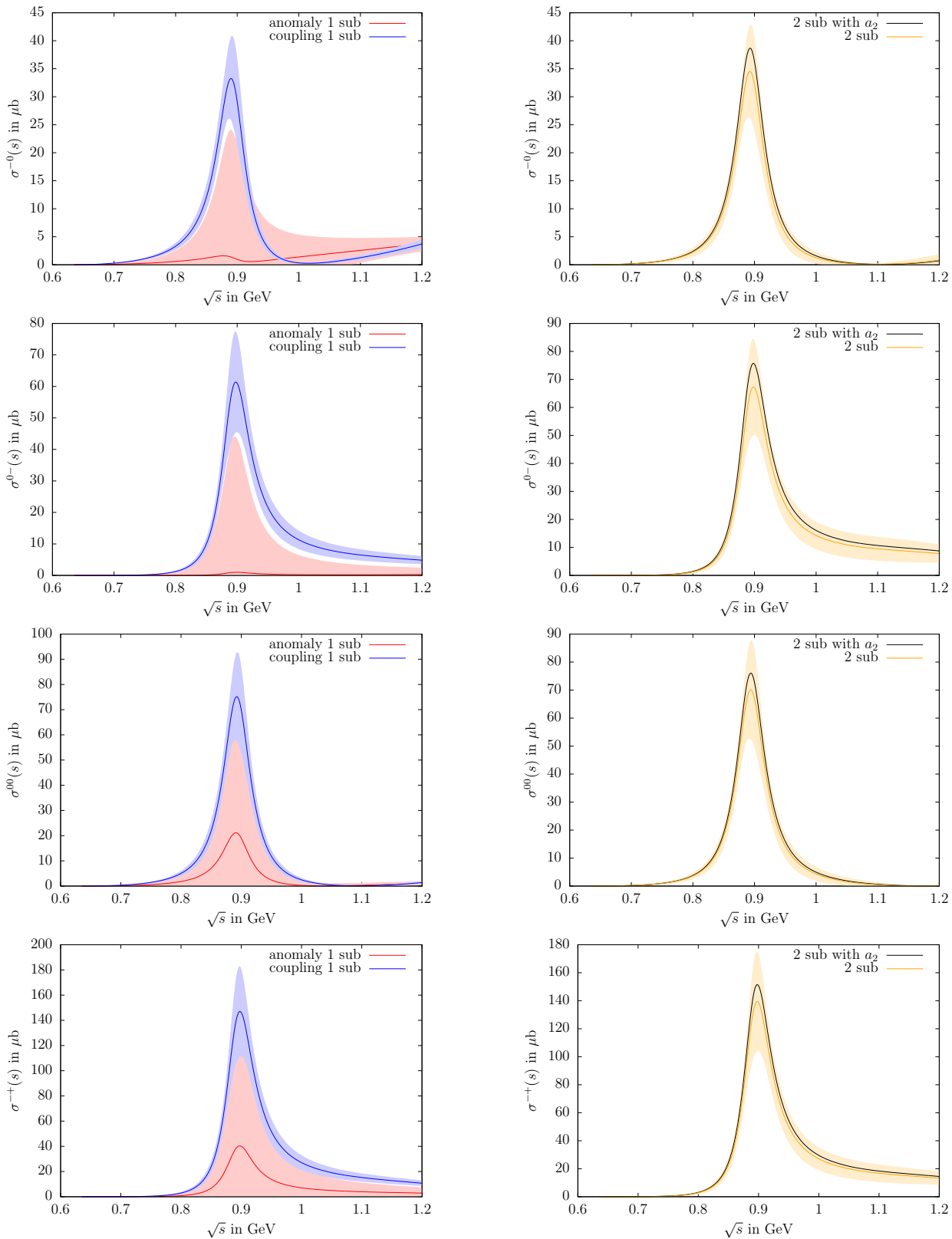

    \fontsize{12pt}{14pt} \selectfont
	\begin{subfigure}{0.49\textwidth}
		\scalebox{0.6}{\input{./plots/cross_sections/F1_1sub_cross.tex}}
	\end{subfigure}
	\hfill
	\begin{subfigure}{0.49\textwidth}
	    \scalebox{0.6}{\input{./plots/cross_sections/F1_2sub_cross.tex}}
	\end{subfigure}
	\hfill
	\begin{subfigure}{0.49\textwidth}
		\scalebox{0.6}{\input{./plots/cross_sections/F2_1sub_cross.tex}}
	\end{subfigure}
	\hfill
	\begin{subfigure}{0.49\textwidth}
	    \scalebox{0.6}{\input{./plots/cross_sections/F2_2sub_cross.tex}}
	\end{subfigure}
	\hfill
	\begin{subfigure}{0.49\textwidth}
		\scalebox{0.6}{\input{./plots/cross_sections/F3_1sub_cross.tex}}
	\end{subfigure}
	\hfill
	\begin{subfigure}{0.49\textwidth}
	    \scalebox{0.6}{\input{./plots/cross_sections/F3_2sub_cross.tex}}
	\end{subfigure}
	\hfill
	\begin{subfigure}{0.49\textwidth}
		\scalebox{0.6}{\input{./plots/cross_sections/F4_1sub_cross.tex}}
	\end{subfigure}
	\hfill
	\begin{subfigure}{0.49\textwidth}
	    \scalebox{0.6}{\input{./plots/cross_sections/F4_2sub_cross.tex}}
	\end{subfigure}
	\caption{Cross section results for $\gamma K^-\to K^- \pi^0$, $\gamma K^-\to \bar{K}^0 \pi^-$, $\gamma \bar{K}^0\to \bar{K}^0 \pi^0$, and $\gamma \bar{K}^0\to K^- \pi^+$ (from top to bottom) including $P$-wave amplitudes. Left panels: minimal subtraction scheme, matched to the chiral anomaly and the $K^*$ radiative couplings separately. Right panels: twice subtracted scheme, matched to anomaly and radiative couplings simultaneously, with and without the $a_2$ contribution. The error bands correspond the propagated error of the real and imaginary parts. 
	}
	\label{fig:cross_sections}
\end{figure*}

Using Eq.~\eqref{eq:crosssection} and the respective partial-wave amplitudes, we can calculate the cross sections for all physical channels, shown in Fig.~\ref{fig:cross_sections}. While the differences between the various channels at low energies, very discernible in the amplitudes, are hardly observable due to the phase space factors---the onset of the visible cross sections only seems to be deferred by about $50\MeV$ for the charge-exchange reactions with their suppressed near-threshold amplitudes---, we see a significant difference between the $\pi^0$ and the charge-exchange channels above the $K^*(892)$, where we predict a strong suppression of the $\pi^0$ production cross sections around $1.1\GeV$.  As we expect $D$-wave corrections to become important only above those energies, such a suppression should be realistically observable in experiments. Note that the order of magnitude for the cross sections is the same as for the related $\gamma\pi\to\pi\pi$ process \cite{Hoferichter:2017ftn}. Furthermore we observe that with incoming neutral kaons, the cross sections are enhanced by about a factor of two compared to their charged-kaon counterparts, while the outgoing-neutral-pion channels are suppressed by again roughly a factor of two in the peak region in comparison to the charge-exchange reactions.

\section{Summary}\label{sec:summary}
In this paper, we have constructed a dispersive representation for the reaction $\gamma K\to K\pi$ that can be measured with kaon beams using the Primakoff mechanism.  Our formalism relies on isospin symmetry and describes all four physical charge channels simultaneously.  We have solved Khuri--Treiman-type equations for the dominant $s$-(and $u$-)channel $P$-wave, based on fixed $t$-channel singularities that are constrained by data and phenomenology as far as possible.  
We have demonstrated that subtraction constants, the free parameters of the theory, can be matched to the low-energy prediction by the chiral anomaly, or the radiative couplings of the $K^*(892)$ resonance, or both.  The dispersive amplitudes provide the correct, model-independent framework to continue data both to the subthreshold region, where it can be matched to chiral perturbation theory, and into the complex-energy plane, where resonance couplings are defined as pole residues on unphysical Riemann sheets.  

\begin{sloppypar}
Options for future theoretical improvement comprise in particular the 
calculation of the next-to-leading-order, or $\Order(p^6)$, corrections to the chiral anomaly for this reaction. 
Furthermore, a reduction of the uncertainty in the $\omega\to K\bar{K}$ coupling, which affects our amplitude representation rather strongly, would be highly desirable; this may be achievable via a refined analysis of the kaon vector form factor.
\end{sloppypar}

Once high-precision, high-statistics experimental data is available, from COMPASS++/AMBER or elsewhere, a simultaneous fit to the two observable charge configurations in $\gamma K^-$ fixes the subtraction constants, from where it is possible to extract the physical quantities of interest. The dispersive representation therefore allows future experiments to determine precise information on the anomaly in a photon--kaon reaction as well as the radiative couplings of the $K^*(892)$ resonance from the complete measured energy range up to $\sqrt{s} \approx 1.2\GeV$.

\begin{acknowledgements}
\bsp
We would like to thank Martin Hoferichter, Tobias Isken, Norbert Kaiser, Malwin Niehus, Stefan Ropertz, Jacobo Ruiz de Elvira, and Mikhail Vysotsky for helpful discussions. 
Partial financial support by the Deutsche Forschungsgemeinschaft (DFG) through funds provided to the Sino--German CRC~110 \enquote{Symmetries and the Emergence of Structure in QCD} 
and by the Bonn--Cologne Graduate School of Physics and Astronomy (BCGS) is gratefully acknowledged.
\esp
\end{acknowledgements}

\begin{appendix}

\section{Kernel functions}\label{app:kernel}
Starting from the definition in Eq.~\eqref{eq:kernelfunc} we can calculate the twice subtracted kernel functions explicitly.
The form of the three kernel functions is given by~\cite{Stamen2020}
\begin{align}
K_2^s(s,s')&=\frac{s^2}{s'^2\left(s'-s\right)}\,,\notag\\
K_2^t(s,s')
&=
\frac{3}{2b}\left[\left(1-x^2\right)Q_0(x)+x\right]-\frac{1}{s'}-\frac{a^{+\Delta}}{s'^2}
\,,\notag\\
K_2^u(s,s')
&=
\frac{3}{2b}\left[\left(1-x^2\right)Q_0(x)+x\right]-\frac{1}{s'}-\frac{a^{-\Delta}}{s'^2}
\,,
\end{align}
where we use the definitions 
\begin{align}
x(s,s')&=\frac{s'-a^{\pm\Delta}(s)}{b(s)}\,,\notag\\
Q_0(x)&=\frac{1}{2}\int_{-1}^{1}\frac{\dd z}{x-z}\,,\notag\\
Q_0(x\pm i\epsilon)&=\frac{1}{2}\log\left|\frac{1+x}{1-x}\right|\mp i\frac{\pi}{2}\theta\left(1-x^2\right)\,.
\end{align}
The once subtracted kernel functions look very similar and read
\begin{align}
K_1^s(s,s')&=\frac{s}{s'\left(s'-s\right)}\,,\notag\\
K_1^t(s,s')
&=
\frac{3}{2b}\left[\left(1-x^2\right)Q_0(x)+x\right]-\frac{1}{s'}
\,,\notag\\
K_1^u(s,s')
&=
\frac{3}{2b}\left[\left(1-x^2\right)Q_0(x)+x\right]-\frac{1}{s'}
\,.
\end{align}
Finally, also the $D$-wave kernel for the $a_2$ contribution can be calculated, which reads
\begin{align}
K_2^{t,D}(s,s')
&=
\left(a^{+\Delta}-s\right) K_2^{t}(s,s') \notag\\
&\quad-\frac{3}{2}x\left(1-x^2\right) Q_0(x)+1-\frac{3(a^{+\Delta})^2}{2b^2} \notag\\
&\quad+\frac{b^2}{5s'^2}+\frac{3a^{+\Delta}s'}{b^2}-\frac{3s'^2}{2b^2}
\,.
\end{align}

\end{appendix}

\bibliographystyle{utphysmod}
\bibliography{Literature}

\providecommand{\href}[2]{#2}\begingroup\raggedright\begin{thebibliography}{10}

\bibitem{Wess:1971yu}
J.~Wess and B.~Zumino,
  \href{http://dx.doi.org/10.1016/0370-2693(71)90582-X}{Phys. Lett. B
  {\bfseries 37}, 95 (1971)}.

\bibitem{Witten:1983tw}
E.~Witten, \href{http://dx.doi.org/10.1016/0550-3213(83)90063-9}{Nucl. Phys. B
  {\bfseries 223}, 422 (1983)}.

\bibitem{Adler:1969gk}
S.~L. Adler, \href{http://dx.doi.org/10.1103/PhysRev.177.2426}{Phys. Rev.
  {\bfseries 177}, 2426 (1969)}.

\bibitem{Bell:1969ts}
J.~Bell and R.~Jackiw, \href{http://dx.doi.org/10.1007/BF02823296}{Nuovo Cim. A
  {\bfseries 60}, 47 (1969)}.

\bibitem{Bardeen:1969md}
W.~A. Bardeen, \href{http://dx.doi.org/10.1103/PhysRev.184.1848}{Phys. Rev.
  {\bfseries 184}, 1848 (1969)}.

\bibitem{Bijnens:1989ff}
J.~Bijnens, A.~Bramon, and F.~Cornet,
  \href{http://dx.doi.org/10.1016/0370-2693(90)91212-T}{Phys. Lett. B
  {\bfseries 237}, 488 (1990)}.

\bibitem{Adler:1971nq}
S.~L. Adler, B.~W. Lee, S.~Treiman, and A.~Zee,
  \href{http://dx.doi.org/10.1103/PhysRevD.4.3497}{Phys. Rev. D {\bfseries 4},
  3497 (1971)}.

\bibitem{Terentev:1971cso}
M.~Terent'ev, \href{http://dx.doi.org/10.1016/0370-2693(72)90171-2}{Phys. Lett.
  B {\bfseries 38}, 419 (1972)}.

\bibitem{Aviv:1971hq}
R.~Aviv and A.~Zee, \href{http://dx.doi.org/10.1103/PhysRevD.5.2372}{Phys. Rev.
  D {\bfseries 5}, 2372 (1972)}.

\bibitem{Primakoff:1951pj}
H.~Primakoff, \href{http://dx.doi.org/10.1103/PhysRev.81.899}{Phys. Rev.
  {\bfseries 81}, 899 (1951)}.

\bibitem{Antipov:1986tp}
Y.~M. Antipov {\em et~al.},
  \href{http://dx.doi.org/10.1103/PhysRevD.36.21}{Phys. Rev. D {\bfseries 36},
  21 (1987)}.

\bibitem{Seyfried2017}
J.~A. Seyfried, Master's thesis, TU München  (2017).

\bibitem{Halprin:1966zz}
A.~Halprin, C.~Andersen, and H.~Primakoff,
  \href{http://dx.doi.org/10.1103/PhysRev.152.1295}{Phys. Rev. {\bfseries 152},
  1295 (1966)}.

\bibitem{Bemporad:1973pa}
C.~Bemporad {\em et~al.},
  \href{http://dx.doi.org/10.1016/0550-3213(73)90496-3}{Nucl. Phys. B
  {\bfseries 51}, 1 (1973)}.

\bibitem{Carithers:1975cg}
W.~Carithers, P.~M{\"u}hlemann, D.~Underwood, and D.~Ryan,
  \href{http://dx.doi.org/10.1103/PhysRevLett.35.349}{Phys. Rev. Lett.
  {\bfseries 35}, 349 (1975)}.

\bibitem{Berg:1980cp}
D.~Berg {\em et~al.},
  \href{http://dx.doi.org/10.1016/0370-2693(81)90380-4}{Phys. Lett. B
  {\bfseries 98}, 119 (1981)}.

\bibitem{Cihangir:1982ze}
S.~Cihangir {\em et~al.} [FERMILAB-MINNESOTA-ROCHESTER Collaboration],
  \href{http://dx.doi.org/10.1016/0370-2693(82)90887-5}{Phys. Lett. B
  {\bfseries 117}, 123 (1982)}.

\bibitem{Chandlee:1983hf}
C.~Chandlee {\em et~al.},
  \href{http://dx.doi.org/10.1103/PhysRevLett.51.168}{Phys. Rev. Lett.
  {\bfseries 51}, 168 (1983)}.

\bibitem{Carlsmith:1985ep}
D.~Carlsmith {\em et~al.},
  \href{http://dx.doi.org/10.1103/PhysRevLett.56.18}{Phys. Rev. Lett.
  {\bfseries 56}, 18 (1986)}.

\bibitem{AlaviHarati:2001ic}
A.~Alavi-Harati {\em et~al.} [KTeV Collaboration],
  \href{http://dx.doi.org/10.1103/PhysRevLett.89.072001}{Phys. Rev. Lett.
  {\bfseries 89}, 072001 (2002)}
  [\href{https://arxiv.org/abs/hep-ex/0110016}{{arXiv:hep-ex/0110016}}].

\bibitem{Burtovoy:2016dyf}
V.~Burtovoy, \href{http://dx.doi.org/10.1134/S1063778815130062}{Phys. Atom.
  Nucl. {\bfseries 78}, 1470 (2015)}.

\bibitem{Burtovoy:2017hbt}
V.~Burtovoy, \href{http://dx.doi.org/10.1134/S1063779617060120}{Phys. Part.
  Nucl. {\bfseries 48}, 932 (2017)}.

\bibitem{Abbon:2014aex}
P.~Abbon {\em et~al.} [COMPASS Collaboration],
  \href{http://dx.doi.org/10.1016/j.nima.2015.01.035}{Nucl. Instrum. Meth. A
  {\bfseries 779}, 69 (2015)}
  [\href{https://arxiv.org/abs/1410.1797}{{arXiv:1410.1797
  [physics.ins-det]}}].

\bibitem{Denisov:2018unj}
B.~Adams {\em et~al.}, 2018.
\newblock \href{https://arxiv.org/abs/1808.00848}{{arXiv:1808.00848 [hep-ex]}}.

\bibitem{Bernhard:2019jqz}
J.~Bernhard {\em et~al.}, \href{http://dx.doi.org/10.1063/5.0008957}{AIP Conf.
  Proc. {\bfseries 2249}, 030035 (2020)}
  [\href{https://arxiv.org/abs/1911.01498}{{arXiv:1911.01498 [hep-ex]}}].

\bibitem{Hoferichter:2012pm}
M.~Hoferichter, B.~Kubis, and D.~Sakkas,
  \href{http://dx.doi.org/10.1103/PhysRevD.86.116009}{Phys. Rev. D {\bfseries
  86}, 116009 (2012)} [\href{https://arxiv.org/abs/1210.6793}{{arXiv:1210.6793
  [hep-ph]}}].

\bibitem{Hoferichter:2017ftn}
M.~Hoferichter, B.~Kubis, and M.~Zanke,
  \href{http://dx.doi.org/10.1103/PhysRevD.96.114016}{Phys. Rev. D {\bfseries
  96}, 114016 (2017)}
  [\href{https://arxiv.org/abs/1710.00824}{{arXiv:1710.00824 [hep-ph]}}].

\bibitem{Briceno:2015dca}
R.~A. Brice\~no, J.~J. Dudek, R.~G. Edwards, C.~J. Shultz, C.~E. Thomas, and
  D.~J. Wilson, \href{http://dx.doi.org/10.1103/PhysRevLett.115.242001}{Phys.
  Rev. Lett. {\bfseries 115}, 242001 (2015)}
  [\href{https://arxiv.org/abs/1507.06622}{{arXiv:1507.06622 [hep-ph]}}].

\bibitem{Briceno:2016kkp}
R.~A. Brice\~no, J.~J. Dudek, R.~G. Edwards, C.~J. Shultz, C.~E. Thomas, and
  D.~J. Wilson, \href{http://dx.doi.org/10.1103/PhysRevD.93.114508}{Phys. Rev.
  D {\bfseries 93}, 114508 (2016)}
  [\href{https://arxiv.org/abs/1604.03530}{{arXiv:1604.03530 [hep-ph]}}].

\bibitem{Alexandrou:2018jbt}
C.~Alexandrou, L.~Leskovec, S.~Meinel, J.~Negele, S.~Paul, M.~Petschlies,
  A.~Pochinsky, G.~Rendon, and S.~Syritsyn,
  \href{http://dx.doi.org/10.1103/PhysRevD.98.074502}{Phys. Rev. D {\bfseries
  98}, 074502 (2018)}
  [\href{https://arxiv.org/abs/1807.08357}{{arXiv:1807.08357 [hep-lat]}}].

\bibitem{Niehus:2019nkl}
M.~Niehus, M.~Hoferichter, and B.~Kubis,
  \href{http://dx.doi.org/10.22323/1.317.0076}{PoS {\bfseries CD2018}, 076
  (2019)} [\href{https://arxiv.org/abs/1902.10150}{{arXiv:1902.10150
  [hep-ph]}}].

\bibitem{Hoferichter:2014vra}
M.~Hoferichter, B.~Kubis, S.~Leupold, F.~Niecknig, and S.~P. Schneider,
  \href{http://dx.doi.org/10.1140/epjc/s10052-014-3180-0}{Eur. Phys. J. C
  {\bfseries 74}, 3180 (2014)}
  [\href{https://arxiv.org/abs/1410.4691}{{arXiv:1410.4691 [hep-ph]}}].

\bibitem{Hoferichter:2018dmo}
M.~Hoferichter, B.-L. Hoid, B.~Kubis, S.~Leupold, and S.~P. Schneider,
  \href{http://dx.doi.org/10.1103/PhysRevLett.121.112002}{Phys. Rev. Lett.
  {\bfseries 121}, 112002 (2018)}
  [\href{https://arxiv.org/abs/1805.01471}{{arXiv:1805.01471 [hep-ph]}}].

\bibitem{Hoferichter:2018kwz}
M.~Hoferichter, B.-L. Hoid, B.~Kubis, S.~Leupold, and S.~P. Schneider,
  \href{http://dx.doi.org/10.1007/JHEP10(2018)141}{JHEP {\bfseries 10}, 141
  (2018)} [\href{https://arxiv.org/abs/1808.04823}{{arXiv:1808.04823
  [hep-ph]}}].

\bibitem{Colangelo:2014pva}
G.~Colangelo, M.~Hoferichter, B.~Kubis, M.~Procura, and P.~Stoffer,
  \href{http://dx.doi.org/10.1016/j.physletb.2014.09.021}{Phys. Lett. B
  {\bfseries 738}, 6 (2014)}
  [\href{https://arxiv.org/abs/1408.2517}{{arXiv:1408.2517 [hep-ph]}}].

\bibitem{Aoyama:2020ynm}
T.~Aoyama {\em et~al.},
  \href{http://dx.doi.org/10.1016/j.physrep.2020.07.006}{Phys. Rept. {\bfseries
  887}, 1 (2020)} [\href{https://arxiv.org/abs/2006.04822}{{arXiv:2006.04822
  [hep-ph]}}].

\bibitem{Berg1983}
D.~M. Berg, PhD thesis, Rochester University  (1983).

\bibitem{GarciaMartin:2010cw}
R.~Garc\'ia-Mart\'in and B.~Moussallam,
  \href{http://dx.doi.org/10.1140/epjc/s10052-010-1471-7}{Eur. Phys. J. C
  {\bfseries 70}, 155 (2010)}
  [\href{https://arxiv.org/abs/1006.5373}{{arXiv:1006.5373 [hep-ph]}}].

\bibitem{Danilkin:2019opj}
I.~Danilkin, O.~Deineka, and M.~Vanderhaeghen,
  \href{http://dx.doi.org/10.1103/PhysRevD.101.054008}{Phys. Rev. D {\bfseries
  101}, 054008 (2020)}
  [\href{https://arxiv.org/abs/1909.04158}{{arXiv:1909.04158 [hep-ph]}}].

\bibitem{Danilkin:2017lyn}
I.~Danilkin, O.~Deineka, and M.~Vanderhaeghen,
  \href{http://dx.doi.org/10.1103/PhysRevD.96.114018}{Phys. Rev. D {\bfseries
  96}, 114018 (2017)}
  [\href{https://arxiv.org/abs/1709.08595}{{arXiv:1709.08595 [hep-ph]}}].

\bibitem{Lu:2020qeo}
J.~Lu and B.~Moussallam,
  \href{http://dx.doi.org/10.1140/epjc/s10052-020-7969-8}{Eur. Phys. J. C
  {\bfseries 80}, 436 (2020)}
  [\href{https://arxiv.org/abs/2002.04441}{{arXiv:2002.04441 [hep-ph]}}].

\bibitem{Guerrero:1997rd}
F.~Guerrero and J.~Prades,
  \href{http://dx.doi.org/10.1016/S0370-2693(97)00656-4}{Phys. Lett. B
  {\bfseries 405}, 341 (1997)}
  [\href{https://arxiv.org/abs/hep-ph/9702303}{{arXiv:hep-ph/9702303}}].

\bibitem{Moinester:2003rb}
M.~Moinester [COMPASS Collaboration], Czech. J. Phys. {\bfseries 53}, B169
  (2003) [\href{https://arxiv.org/abs/hep-ex/0301024}{{arXiv:hep-ex/0301024}}].

\bibitem{Vysotsky:2015mks}
M.~Vysotsky and E.~Zhemchugov,
  \href{http://dx.doi.org/10.1103/PhysRevD.93.094029}{Phys. Rev. D {\bfseries
  93}, 094029 (2016)}
  [\href{https://arxiv.org/abs/1512.04438}{{arXiv:1512.04438 [hep-ph]}}],
  [Erratum: Phys.Rev.D 94, 019901 (2016)].

\bibitem{Ebertshaeuser2001}
T.~Ebertshäuser, PhD thesis, Mainz University  (2001).

\bibitem{Hacker2008}
C.~Hacker, PhD thesis, Mainz University  (2008).

\bibitem{Khuri:1960zz}
N.~Khuri and S.~Treiman,
  \href{http://dx.doi.org/10.1103/PhysRev.119.1115}{Phys. Rev. {\bfseries 119},
  1115 (1960)}.

\bibitem{Buettiker:2003pp}
P.~B{\"u}ttiker, S.~Descotes-Genon, and B.~Moussallam,
  \href{http://dx.doi.org/10.1140/epjc/s2004-01591-1}{Eur. Phys. J. C
  {\bfseries 33}, 409 (2004)}
  [\href{https://arxiv.org/abs/hep-ph/0310283}{{arXiv:hep-ph/0310283}}].

\bibitem{Pelaez:2016tgi}
J.~Pel\'aez and A.~Rodas,
  \href{http://dx.doi.org/10.1103/PhysRevD.93.074025}{Phys. Rev. D {\bfseries
  93}, 074025 (2016)}
  [\href{https://arxiv.org/abs/1602.08404}{{arXiv:1602.08404 [hep-ph]}}].

\bibitem{Pelaez:2020gnd}
J.~Pel\'aez and A.~Rodas, 2020.
\newblock \href{https://arxiv.org/abs/2010.11222}{{arXiv:2010.11222 [hep-ph]}}.

\bibitem{Mandelstam:1958xc}
S.~Mandelstam, \href{http://dx.doi.org/10.1103/PhysRev.112.1344}{Phys. Rev.
  {\bfseries 112}, 1344 (1958)}.

\bibitem{Jacob:1959at}
M.~Jacob and G.~Wick,
  \href{http://dx.doi.org/10.1016/0003-4916(59)90051-X}{Annals Phys. {\bfseries
  7}, 404 (1959)}.

\bibitem{Chew:1957tf}
G.~Chew, M.~Goldberger, F.~Low, and Y.~Nambu,
  \href{http://dx.doi.org/10.1103/PhysRev.106.1345}{Phys. Rev. {\bfseries 106},
  1345 (1957)}.

\bibitem{Tiator:1990ck}
L.~Tiator and D.~Drechsel,
  \href{http://dx.doi.org/10.1016/0375-9474(90)90516-O}{Nucl. Phys. A
  {\bfseries 508}, 541C (1990)}.

\bibitem{Drechsel:1992pn}
D.~Drechsel and L.~Tiator,
  \href{http://dx.doi.org/10.1088/0954-3899/18/3/004}{J. Phys. G {\bfseries
  18}, 449 (1992)}.

\bibitem{Hanstein:1996bd}
O.~Hanstein, D.~Drechsel, and L.~Tiator,
  \href{http://dx.doi.org/10.1016/S0370-2693(97)00272-4}{Phys. Lett. B
  {\bfseries 399}, 13 (1997)}
  [\href{https://arxiv.org/abs/nucl-th/9612057}{{arXiv:nucl-th/9612057}}].

\bibitem{Hanstein:1997tp}
O.~Hanstein, D.~Drechsel, and L.~Tiator,
  \href{http://dx.doi.org/10.1016/S0375-9474(98)00818-5}{Nucl. Phys. A
  {\bfseries 632}, 561 (1998)}
  [\href{https://arxiv.org/abs/nucl-th/9709067}{{arXiv:nucl-th/9709067}}].

\bibitem{Pasquini:2006yi}
B.~Pasquini, D.~Drechsel, and L.~Tiator,
  \href{http://dx.doi.org/10.1140/epja/i2006-10004-2}{Eur. Phys. J. A
  {\bfseries 27}, 231 (2006)}
  [\href{https://arxiv.org/abs/nucl-th/0603006}{{arXiv:nucl-th/0603006}}].

\bibitem{Stern:1993rg}
J.~Stern, H.~Sazdjian, and N.~Fuchs,
  \href{http://dx.doi.org/10.1103/PhysRevD.47.3814}{Phys. Rev. D {\bfseries
  47}, 3814 (1993)}
  [\href{https://arxiv.org/abs/hep-ph/9301244}{{arXiv:hep-ph/9301244}}].

\bibitem{Knecht:1995tr}
M.~Knecht, B.~Moussallam, J.~Stern, and N.~Fuchs,
  \href{http://dx.doi.org/10.1016/0550-3213(95)00515-3}{Nucl. Phys. B
  {\bfseries 457}, 513 (1995)}
  [\href{https://arxiv.org/abs/hep-ph/9507319}{{arXiv:hep-ph/9507319}}].

\bibitem{Zdrahal:2008bd}
M.~Zdr\'ahal and J.~Novotn\'y,
  \href{http://dx.doi.org/10.1103/PhysRevD.78.116016}{Phys. Rev. D {\bfseries
  78}, 116016 (2008)} [\href{https://arxiv.org/abs/0806.4529}{{arXiv:0806.4529
  [hep-ph]}}].

\bibitem{Dax2020}
M.~Dax, Master's thesis, Bonn University  (2020).

\bibitem{Moussallam:2007qc}
B.~Moussallam, \href{http://dx.doi.org/10.1140/epjc/s10052-007-0464-7}{Eur.
  Phys. J. C {\bfseries 53}, 401 (2008)}
  [\href{https://arxiv.org/abs/0710.0548}{{arXiv:0710.0548 [hep-ph]}}].

\bibitem{Niecknig:2012sj}
F.~Niecknig, B.~Kubis, and S.~P. Schneider,
  \href{http://dx.doi.org/10.1140/epjc/s10052-012-2014-1}{Eur. Phys. J. C
  {\bfseries 72}, 2014 (2012)}
  [\href{https://arxiv.org/abs/1203.2501}{{arXiv:1203.2501 [hep-ph]}}].

\bibitem{Albaladejo:2018gif}
M.~Albaladejo, N.~Sherrill, C.~Fern\'andez-Ram\'\i{}rez, A.~Jackura,
  V.~Mathieu, M.~Mikhasenko, J.~Nys, A.~Pilloni, and A.~Szczepaniak [JPAC
  Collaboration], \href{http://dx.doi.org/10.1140/epjc/s10052-018-6045-0}{Eur.
  Phys. J. C {\bfseries 78}, 574 (2018)}
  [\href{https://arxiv.org/abs/1803.06027}{{arXiv:1803.06027 [hep-ph]}}].

\bibitem{Colangelo:2018jxw}
G.~Colangelo, S.~Lanz, H.~Leutwyler, and E.~Passemar,
  \href{http://dx.doi.org/10.1140/epjc/s10052-018-6377-9}{Eur. Phys. J. C
  {\bfseries 78}, 947 (2018)}
  [\href{https://arxiv.org/abs/1807.11937}{{arXiv:1807.11937 [hep-ph]}}].

\bibitem{Isken:2017dkw}
T.~Isken, B.~Kubis, S.~P. Schneider, and P.~Stoffer,
  \href{http://dx.doi.org/10.1140/epjc/s10052-017-5024-1}{Eur. Phys. J. C
  {\bfseries 77}, 489 (2017)}
  [\href{https://arxiv.org/abs/1705.04339}{{arXiv:1705.04339 [hep-ph]}}].

\bibitem{Niecknig:2015ija}
F.~Niecknig and B.~Kubis, \href{http://dx.doi.org/10.1007/JHEP10(2015)142}{JHEP
  {\bfseries 10}, 142 (2015)}
  [\href{https://arxiv.org/abs/1509.03188}{{arXiv:1509.03188 [hep-ph]}}].

\bibitem{Niecknig:2017ylb}
F.~Niecknig and B.~Kubis,
  \href{http://dx.doi.org/10.1016/j.physletb.2018.03.048}{Phys. Lett. B
  {\bfseries 780}, 471 (2018)}
  [\href{https://arxiv.org/abs/1708.00446}{{arXiv:1708.00446 [hep-ph]}}].

\bibitem{Hoferichter:2011wk}
M.~Hoferichter, D.~R. Phillips, and C.~Schat,
  \href{http://dx.doi.org/10.1140/epjc/s10052-011-1743-x}{Eur. Phys. J. C
  {\bfseries 71}, 1743 (2011)}
  [\href{https://arxiv.org/abs/1106.4147}{{arXiv:1106.4147 [hep-ph]}}].

\bibitem{Kubis:2015sga}
B.~Kubis and J.~Plenter,
  \href{http://dx.doi.org/10.1140/epjc/s10052-015-3495-5}{Eur. Phys. J. C
  {\bfseries 75}, 283 (2015)}
  [\href{https://arxiv.org/abs/1504.02588}{{arXiv:1504.02588 [hep-ph]}}].

\bibitem{Klingl:1996by}
F.~Klingl, N.~Kaiser, and W.~Weise,
  \href{http://dx.doi.org/10.1007/s002180050167}{Z. Phys. A {\bfseries 356},
  193 (1996)}
  [\href{https://arxiv.org/abs/hep-ph/9607431}{{arXiv:hep-ph/9607431}}].

\bibitem{Terschluesen:2010ik}
C.~Terschl{\"u}sen and S.~Leupold,
  \href{http://dx.doi.org/10.1016/j.physletb.2010.06.033}{Phys. Lett. B
  {\bfseries 691}, 191 (2010)}
  [\href{https://arxiv.org/abs/1003.1030}{{arXiv:1003.1030 [hep-ph]}}].

\bibitem{Okubo:1963fa}
S.~Okubo, \href{http://dx.doi.org/10.1016/S0375-9601(63)92548-9}{Phys. Lett.
  {\bfseries 5}, 165 (1963)}.

\bibitem{Zweig1964}
G.~Zweig. Unpublished, 1964.

\bibitem{Iizuka:1966wu}
J.~Iizuka, K.~Okada, and O.~Shito,
  \href{http://dx.doi.org/10.1143/PTP.35.1061}{Prog. Theor. Phys. {\bfseries
  35}, 1061 (1966)}.

\bibitem{Zyla:2020zbs}
P.~Zyla {\em et~al.} [Particle Data Group],
  \href{http://dx.doi.org/10.1093/ptep/ptaa104}{PTEP {\bfseries 2020}, 083C01
  (2020)}.

\bibitem{GarciaMartin:2011jx}
R.~Garc\'ia-Mart\'in, R.~Kami\'nski, J.~Pel\'aez, and J.~Ruiz~de Elvira,
  \href{http://dx.doi.org/10.1103/PhysRevLett.107.072001}{Phys. Rev. Lett.
  {\bfseries 107}, 072001 (2011)}
  [\href{https://arxiv.org/abs/1107.1635}{{arXiv:1107.1635 [hep-ph]}}].

\bibitem{Beloborodov:2019fmw}
K.~Beloborodov, V.~Druzhinin, and S.~Serednyakov,
  \href{http://dx.doi.org/10.1134/S1063776119080016}{J. Exp. Theor. Phys.
  {\bfseries 129}, 386 (2019)}.

\bibitem{Pelaez:2018qny}
J.~Pel\'aez and A.~Rodas,
  \href{http://dx.doi.org/10.1140/epjc/s10052-018-6296-9}{Eur. Phys. J. C
  {\bfseries 78}, 897 (2018)}
  [\href{https://arxiv.org/abs/1807.04543}{{arXiv:1807.04543 [hep-ph]}}].

\bibitem{Ecker:2007us}
G.~Ecker and C.~Zauner,
  \href{http://dx.doi.org/10.1140/epjc/s10052-007-0372-x}{Eur. Phys. J. C
  {\bfseries 52}, 315 (2007)}
  [\href{https://arxiv.org/abs/0705.0624}{{arXiv:0705.0624 [hep-ph]}}].

\bibitem{Watson:1954uc}
K.~M. Watson, \href{http://dx.doi.org/10.1103/PhysRev.95.228}{Phys. Rev.
  {\bfseries 95}, 228 (1954)}.

\bibitem{Bar:2001qk}
O.~B{\"a}r and U.~Wiese,
  \href{http://dx.doi.org/10.1016/S0550-3213(01)00288-7}{Nucl. Phys. B
  {\bfseries 609}, 225 (2001)}
  [\href{https://arxiv.org/abs/hep-ph/0105258}{{arXiv:hep-ph/0105258}}].

\bibitem{Stamen2020}
D.~Stamen, Master's thesis, Bonn University  (2020).

\bibitem{Pelaez:2016klv}
J.~Pel\'aez, A.~Rodas, and J.~Ruiz~de Elvira,
  \href{http://dx.doi.org/10.1140/epjc/s10052-017-4668-1}{Eur. Phys. J. C
  {\bfseries 77}, 91 (2017)}
  [\href{https://arxiv.org/abs/1612.07966}{{arXiv:1612.07966 [hep-ph]}}].

\bibitem{Elvira2019}
J.~Ruiz~de Elvira. Private communication, 2019.

\bibitem{Ellis:2016jkw}
J.~Ellis, \href{http://dx.doi.org/10.1016/j.cpc.2016.08.019}{Comput. Phys.
  Commun. {\bfseries 210}, 103 (2017)}
  [\href{https://arxiv.org/abs/1601.05437}{{arXiv:1601.05437 [hep-ph]}}].

\end{thebibliography}\endgroup

\end{document}